\documentclass[fleqn,usenatbib]{mnras}

\usepackage{newtxtext,newtxmath}

\usepackage[T1]{fontenc}
\usepackage{ae,aecompl}
\usepackage{nicefrac}

\usepackage{graphicx}	
\usepackage{amsmath}	
\usepackage{lscape} 
\usepackage{xcolor}

\newcommand{\tess}{\emph{TESS}}
\newcommand{\pyaneti}{\texttt{pyaneti}}

\title[Planet Hunters TESS]{Planet Hunters TESS II: Findings from the first two years of \tess}

\author[Eisner et al.]{
N. L. Eisner,$^{1}$\thanks{E-mail: nora.eisner@physics.ox.ac.uk}
O. Barrag\'an,$^{1}$
C. Lintott,$^{1}$
S. Aigrain,$^{1}$
B. Nicholson,$^{1,2}$
\newauthor
T. S. Boyajian,$^{3}$
S. Howell,$^{4}$
C. Johnston,$^{5,6}$
B. Lakeland,$^{1}$
G. Miller,$^{1}$
\newauthor
A. McMaster,$^{1}$
H. Parviainen,$^{7}$
E. J. Safron,$^{3}$
M. E. Schwamb,$^{8,9}$
L. Trouille,$^{10}$
\newauthor
S. Vaughan,$^{1}$
N. Zicher,$^{1}$
C. Allen, $^{1}$
S. Allen,$^{10}$
M. Bouslog, $^{10}$
C. Johnson,$^{10}$
\newauthor
M. N. Simon,$^{10}$
Z. Wolfenbarger,$^{10}$
E.M.L. Baeten,$^{11}$
D.M. Bundy,$^{11}$
T. Hoffman$^{11}$\thanks{The complete list of Planet Hunters TESS citizen scientists who contributed to this work can be found at \url{https://www.zooniverse.org/projects/nora-dot-eisner/planet-hunters-tess/about/results}. }
\\
$^{1}$Sub-department of Astrophysics, University of Oxford, Keble Rd, Oxford, United Kingdom\\
$^{2}$University of Southern Queensland, Centre for Astrophysics, Toowoomba, Australia\\
$^{3}$Department of Physics and Astronomy, Louisiana State University, Baton Rouge, LA 70803 USA\\
$^{4}$NASA Ames Research Center, Moffett Field, CA 94035, USA\\
$^{5}$Institute of Astronomy, KU Leuven, Celestijnenlaan 200D, 3001 Leuven, Belgium\\
$^{6}$Department of Astrophysics, IMAPP, Radboud University Nijmegen, P. O. Box 9010, 6500 GL Nijmegen, the Netherlands\\
$^{7}$Instituto de Astrofisica de Canarias (IAC), Via Lactea s/n, 38205 La Laguna, Tenerife, Spain\\
$^{8}$Gemini Observatory, Northern Operations Center, 670 North A'ohoku Place, Hilo, HI 96720, USA\\
$^{9}$Astrophysics Research Centre, Queen's University Belfast, Belfast BT7 1NN, UK\\
$^{10}$The Adler Planetarium, Zooniverse, Northwestern University, Chicago, IL\\
$^{11}$Citizen Scientist, Zooniverse c/o University of Oxford,
Keble Road, Oxford OX1 3RH, UK}
\date{Submitted on 5 October 2020; revised on 17 November 2020; accepted for publication by MNRAS on 23 November 2020.}

\pubyear{2020}

\begin{document}
\label{firstpage}
\pagerange{\pageref{firstpage}--\pageref{lastpage}}
\maketitle

\begin{abstract}
We present the results from the first two years of the Planet Hunters TESS citizen science project, which identifies planet candidates in the \tess\ data by engaging members of the general public. \textcolor{black}{Over 22,000 citizen scientists from around the world visually inspected the first 26 Sectors of TESS data in order to help identify transit-like signals. We use a clustering algorithm to combine these classifications into a ranked list of events for each sector, the top 500 of which are then visually vetted by the science team. We assess the detection efficiency of this methodology by comparing our results to the list of TESS Objects of Interest (TOIs) and show that we recover~85 \% of the TOIs with radii greater than 4 $_\oplus$ and 51~\% of those with radii between 3 and 4 R$_\oplus$.} Additionally, we present our 90 most promising planet candidates \textcolor{black}{that had not previously been identified by other teams}, 73 of which exhibit only a single transit event in the \tess\ light curve, and outline our efforts to follow these candidates up using ground-based observatories. Finally, we present noteworthy stellar systems that were identified through the Planet Hunters TESS project.
\end{abstract}




\section{Introduction}

Since the first unambiguous discovery of an exoplanet in 1995 \citep[][]{Mayor1995} over 4,000 more have been confirmed. Studies of their characteristics have unveiled an extremely wide range of planetary properties in terms of planetary mass, size, system architecture and orbital periods, greatly revolutionising our understanding of how these bodies form and evolve.

The transit method, whereby we observe a temporary decrease in the brightness of a star due to a planet passing in front of its host star, is to date the most successful method for planet detection, having discovered over 75\% of the planets listed on the NASA Exoplanet Archive\footnote{\url{https://exoplanetarchive.ipac.caltech.edu/}}. It yields a wealth of information including planet radius, orbital period, system orientation and potentially even atmospheric composition. Furthermore, when combined with Radial Velocity \citep[RV; e.g.,][]{Mayor1995, Marcy1997} observations, which yield the planetary mass, we can infer planet densities, and thus their internal bulk compositions. Other indirect detection methods include radio pulsar timing \citep[e.g.,][]{Wolszczan1992} and microlensing \citep[e.g.,][]{Gaudi2012}.


The \textit{Transiting Exoplanet Survey Satellite} mission \citep[\protect\tess;][]{ricker15} is currently in its extended mission, searching for transiting planets orbiting bright ($V < 11$\,mag) nearby stars. Over the course of the two year nominal mission, \tess\ monitored around 85 per cent of the sky, split up into 26 rectangular sectors of 96 $\times$ 24 deg each (13 per hemisphere). Each sector is monitored for $\approx$ 27.4 continuous days, measuring the brightness of $\approx$ 20,000 pre-selected stars every two minutes. In addition to these short cadence (SC) observations, the \tess\ mission provides Full Frame Images (FFI) that span across all pixels of all CCDs and are taken at a cadence of 30 minutes. While most of the targets ($\sim$ 63 per cent) will be observed for $\approx$ 27.4 continuous days, around $\sim$ 2 per cent of the targets at the ecliptic poles are located in the `continuous viewing zones' and will be continuously monitored for $\sim$ 356 days.

Stars themselves are extremely complex, with phenomena ranging from outbursts to long and short term variability and oscillations, which manifest themselves in the light curves. These signals, as well as systematic effects and artifacts introduced by the telescope and instruments, mean that standard periodic search methods, such as the Box-Least-Squared method \citep{bls2002} can struggle to identify certain transit events, especially if the observed signal is dominated by natural stellar variability. Standard detection pipelines also tend to bias the detection of short period planets, as they typically require a minimum of two transit events in order to gain the signal-to-noise ratio (SNR) required for detection.

One of the prime science goals of the \tess\ mission is to further our understanding of the overall planet population, an active area of research that is strongly affected by observational and detection biases. In order for exoplanet population studies to be able to draw meaningful conclusions, they require a certain level of completeness in the sample of known exoplanets as well as a robust sample of validated planets spanning a wide range of parameter space. \textcolor{black}{Due to this, we independently search the \tess\ light curves for transiting planets via visual vetting in order to detect candidates that were either intentionally ignored by the main \tess\ pipelines, which require at least two transits for a detection, missed because of stellar variability or instrumental artefacts, or were identified but subsequently erroneously discounted at the vetting stage, usually because the period found by the pipeline was incorrect. These candidates can help populate under-explored regions of parameter space and will, for example, benefit the study of planet occurrence rates around different stellar types as well as inform theories of physical processes involved with the formation and evolution of different types of exoplanets.}

Human brains excel in activities related to pattern recognition, making the task of identifying transiting events in light curves, even when the pattern is in the midst of a strong varying signal, ideally suited for visual vetting. Early citizen science projects, such as Planet Hunters \citep[PH;][]{fischer12} and Exoplanet Explorers \citep{Christiansen2018}, successfully harnessed the analytic power of a large number of volunteers and made substantial contributions to the field of exoplanet discoveries. The PH project, for example, showed that human vetting has a higher detection efficiency than automated detection algorithms for certain types of transits. In particular, they showed that citizen science can outperform on the detection of single (long-period) transits \citep[e.g.,][]{wang13, schmitt14a}, aperiodic transits \citep[e.g. circumbinary planets;][]{schwamb13} and planets around variable stars \citep[e.g., young systems,][]{fischer12}. Both PH and Exoplanet Explorers, which are hosted by the world's largest citizen science platform Zooniverse \citep{lintott08}, ensured easy access to \textit{Kepler} and \textit{K2} data by making them publicly available online in an immediately accessible graphical format that is easy to understand for non-specialists. The popularity of these projects is reflected in the number of participants, with PH attracting 144,466 volunteers from 137 different countries over 9 years of the project being active.

Following the end of the \textit{Kepler} mission and the launch of the \tess\ satellite in 2018, PH was relaunched as the new citizen science project \textit{Planet Hunters TESS} (PHT) \footnote{\url{www.planethunters.org}}, with the aim of identifying transit events in the \tess\ data that were \textcolor{black}{intentionally ignored or missed} by the main \tess\ pipelines. \textcolor{black}{Such a search complements other methods methods via its sensitivity to single-transit, and, therefore, longer period planets. Additionally, other dedicated non-citizen science based methods are also employed to look for single transit candidates \citep[see e.g., the Bayesian transit fitting method by ][]{Gill2020, Osborn2016}}.

Citizen science transit searches specialise in finding the rare events that the standard detection pipelines miss, however, these results are of limited use without an indication of the completeness of the search. Addressing the problem of completeness was therefore one of our highest priorities while designing PHT as discussed throughout this paper. 

The layout of the remainder of the paper will be as follows. An overview of the Planet Hunters TESS project is found in Section~\ref{sec:PHT}, followed by an in depth description of how the project identifies planet candidates in Section~\ref{sec:method}. The recovery efficiency of the citizen science approach is assessed in Section~\ref{sec:recovery_efficiency}, followed by a description of the in-depth vetting of candidates and ground based-follow up efforts in Section~\ref{sec:vetting} and \ref{sec:follow_up}, respectively. Planet Candidates and noteworthy systems identified by Planet Hunters TESS are outlined in Section~\ref{sec:PHT_canidates}, followed by a discussion of the results in Section~\ref{sec:condlusion}.

\section{Planet Hunters TESS}
\label{sec:PHT}

\begin{figure*}
    \centering
    \includegraphics[width=\textwidth]{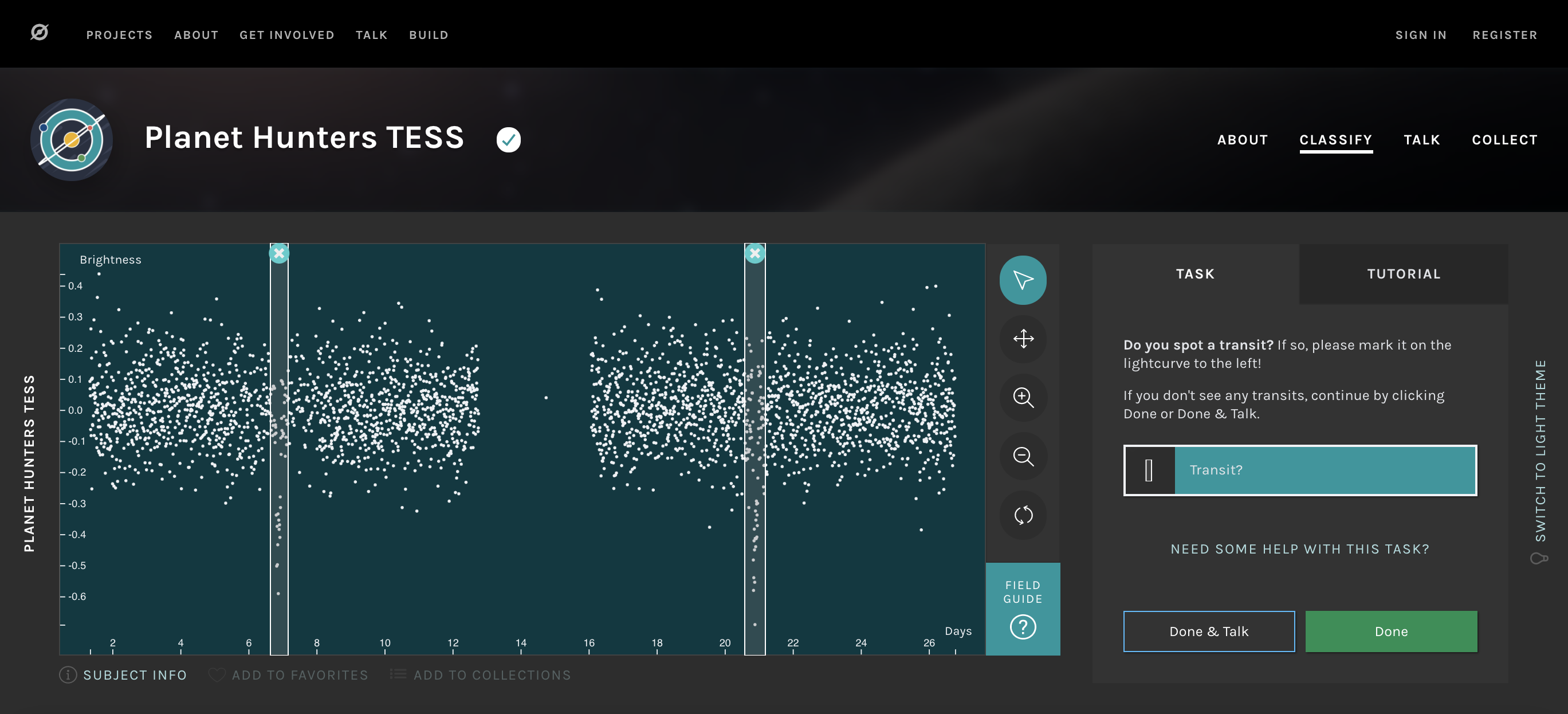}
    \caption{
    PHT user interface showing a simulated light curve. The transit events are highlighted with white partially-transparent columns that are drawn on using the mouse. Stellar information on the target star is available by clicking on `subject info' below the light curve.} 
    \label{fig:interface}
\end{figure*}

The PHT project works by displaying \tess\ light curves (Figure~\ref{fig:interface}), and asking volunteers to identify transit-like signals. Only the two-minute cadence targets, which are produced by the \tess\ pipeline at the Science Processing Operations Center \citep[SPOC,][]{Jenkins2018} and made publicly available by the Mikulski Archive for Space Telescopes (MAST)\footnote{\url{http://archive.stsci.edu/tess/}}, are searched by PHT. First-time visitors to the PHT site, or returning visitors who have not logged in are prompted to look through a short tutorial, which briefly explains the main aim of the project and shows examples of transit events and other stellar phenomena. Scientific explanation of the project can be found elsewhere on the site in the `field guide' and on the project's `About' page.

After viewing the tutorial, volunteers are ready to participate in the project and are presented with \tess\ light curves (known as `subjects') that need to be classified. The project was designed to be as simple as possible and therefore only asks one question: \textit{`Do you see a transit?}'. Users identify transit-like events, and the time of their occurrence, by drawing a column over the event using the mouse button, as shown in Figure~\ref{fig:interface}. There is no limit on the number of transit-like events that can be marked in a light curve. No markings indicate that there are no transit-like events present in the light curve. Once the subject has been analysed, users submit their classification and continue to view the next light curve by clicking `Done'. 

Alongside each light curve, users are offered information on the stellar properties of the target, such as the radius, effective temperature and magnitude (subject to availability, see \cite{Stassun18}). However, in order to reduce biases in the classifications, the TESS Input Catalog (TIC) ID of the target star is not provided until after the subject classification has been submitted.

In addition to classifying the data, users are given the option to comment on light curves via the `Talk' discussion forum. Each light curve has its own discussion page to allow volunteers to discuss and comment, as well as to `tag' light curves using searchable hashtags, and to bring promising candidates to the attention of other users and the research team. The talk discussion forums complement the main PHT analysis and have been shown to yield interesting objects which may be challenging to detect using automated algorithms \citep[e.g.,][]{eisner2019RN}. Unlike in the initial PH project, there are no questions in the main interface regarding stellar variability, however, volunteers are encouraged to mention astrophysical phenomenon or \textit{unusual} features, such as eclipsing binaries or stellar flares, using the `Talk' discussion forum.  

The subject TIC IDs are revealed on the subject discussion pages, allowing volunteers to carry out further analysis on specific targets of interest and to report and discuss their findings. This is extremely valuable for both other volunteers and the PHT science team, as it can speed up the process of identifying candidates as well as rule out false positives in a fast and effective manner. 

Since the launch of PHT on 6 December 2018, there has been one significant makeover to the user interface. The initial PHT user interface (UI1), which was used for sectors 1 through 9, split the \tess\ light curves up into either three or four chunks (depending on the data gaps in each sector) which lasted around seven days each. This allowed for a more `zoomed' in view of the data, making it easier to identify transit-like events than when the full $\sim$ 30 day light curves were shown. The results from a PHT beta project, which displayed only simulated data, showed that a more zoomed in view of the light curve was likely to yield a higher transit recovery rate.

The updated, and current, user interface (UI2) allows users to manually zoom in on the x-axis (time) of the data. Due to this additional feature, each target has been displayed as a single light curve as of Sector 10. In order to verify that the changes in interface did not affect our findings, all of the Sector 9 subjects were classified using both UI1 and UI2. We saw no significant change in the number of candidates recovered (see Section~\ref{sec:recovery_efficiency} for a description of how we quantified detection efficiency).


\subsection{Simulated Data}
\label{subsec:sims} 

In addition to the real data, volunteers are shown simulated light curves, which are generated by randomly injecting simulated transit signals, provided by the SPOC pipeline \citep[][]{Jenkins2018}, into real \tess\ light curves. The simulated data play an important role in assessing the sensitivity of the project, training the users and providing immediate feedback, and to gauge the relative abilities of individual users (see Sec~\ref{subsec:weighting}). 

We calculate a signal to noise ratio (SNR) of the injected signal by dividing the injected transit depth by the Root Mean Square Combined Differential Photometric Precision (RMS CDPP) of the light curve on 0.5-, 1- or 2-hr time scales (whichever is closest to the duration of the injected transit signal). Only simulations with a SNR greater than 7 in UI1 and greater than 4 for UI2 are shown to volunteers.

Simulated light curves are randomly shown to the volunteers and classified in the exact same manner as the real data. The user is always notified after a simulated light curve has been classified and given feedback as to whether the injected signal was correctly identified or not. For each sector, we generate between one and two thousand simulated light curves, using the real data from that sector in order to ensure that the sector specific systematic effects and data gaps of the simulated data do not differ from the real data. The rate at which a volunteer is shown simulated light curves decreases from an initial rate of 30 per cent for the first 10 classifications, down to a rate of 1 per cent by the time that the user has classified 100 light curves.

\section{Identifying Candidates}
\label{sec:method}

Each subject is seen by multiple volunteers, before it is `retired' from the site, and the classifications are combined (see Section~\ref{subsec:DBscan}) in order to assess the likelihood of a transit event. For sectors 1 through 9, the subjects were retired after 8 classifications if the first 8 volunteers who saw the light curves did not mark any transit events, after 10 classifications if the first 10 volunteers all marked a transit event and after 15 classifications if there was not complete consensus amongst the users. As of Sector 9 with UI2, all subjects were classified by 15 volunteers, regardless of whether or not any transit-like events were marked. Sector 9, which was classified with both UI1 and UI2, was also classified with both retirement rules.

\begin{figure}
    \centering
    \includegraphics[width=0.5\textwidth]{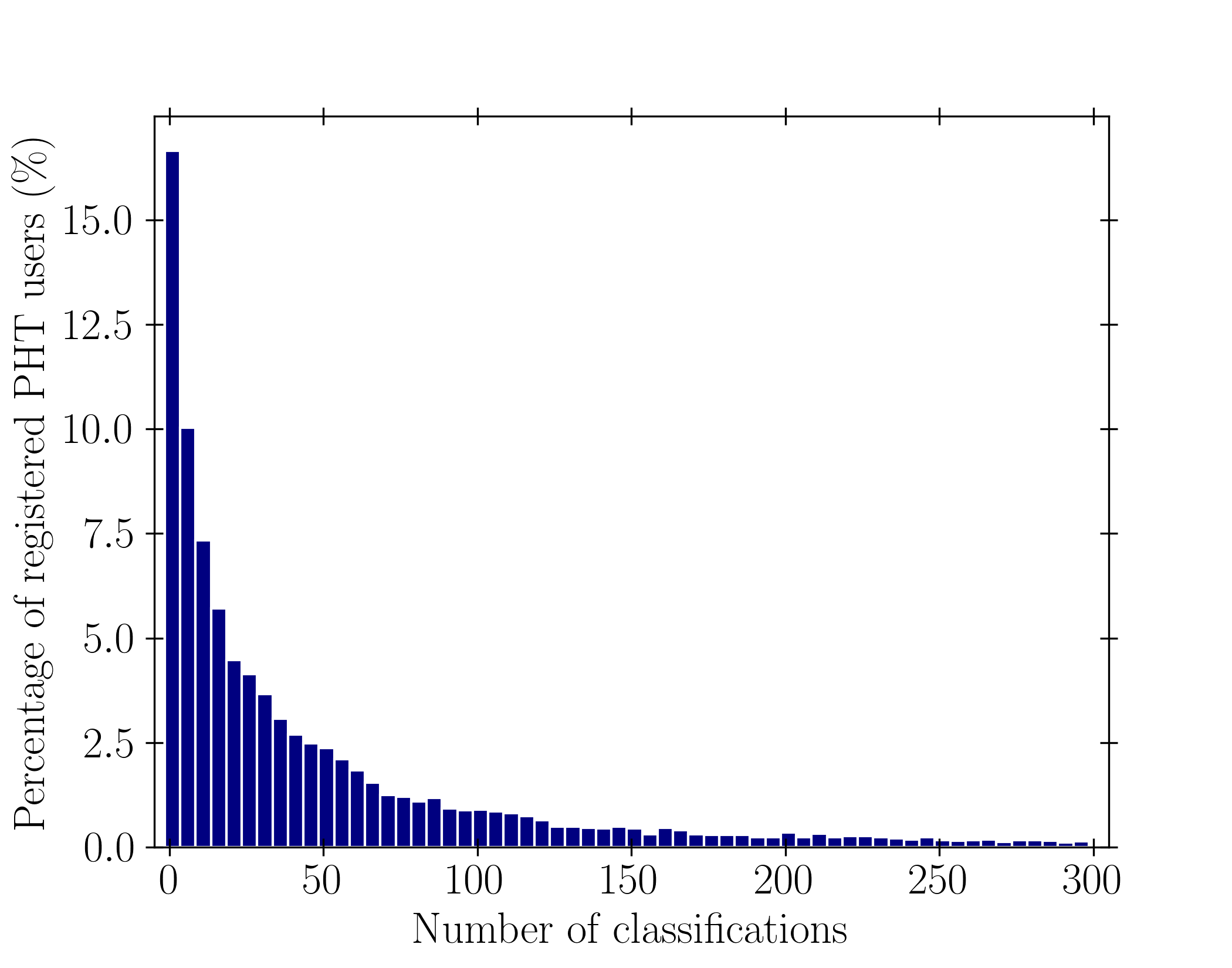}
    \caption{
    The distribution of the number of classifications by the registered volunteers, using a bin size of 5 from 0 to 300 classifications. A total of 11.8 per cent of the registered volunteers completed more than 300 classifications.} 
    \label{fig:user_count}
\end{figure}

There were a total of 12,617,038 individual classifications completed across the project on the nominal mission data. 95.4 per cent of these classifications were made by 22,341 registered volunteers, with the rest made by unregistered volunteers. Around 25 per cent of the registered volunteers complete more than 100 classifications, 11.8 per cent more than 300, 8.4 per cent more than 500, 5.4 per cent more than 1000 and 1.1 per cent more than 10,000. The registered volunteers completed a mean and median of 647 and 33 classifications, respectively. Figure~\ref{fig:user_count} shows the distribution in user effort for logged in users who made between 0 and 300 classifications. 

The distribution in the number of classifications made by the registered volunteers is assessed using the Gini coefficient, which ranges from 0 (equal contributions from all users) to 1 (large disparity in the contributions). The Gini coefficients for individual sectors ranges from 0.84 to 0.91 with a mean of 0.87, while the Gini coefficient for the overall project (all of the sectors combined) is 0.94. The mean Gini coefficient among other astronomy Zooniverse projects lies at 0.82 \citep{spiers2019}. We note that the only other Zooniverse project with an equally high Gini coefficient as PHT is \textit{Supernova Hunters}, a project which, similarly to PHT and unlike most other Zooniverse projects, has periodic data releases that are accompanied by an e-newsletter sent to all project volunteers. Periodic e-newsletters have the effect of promoting the project to both regularly and irregularly participating volunteers, who may only complete a couple of classifications as they explore the task, as well as to returning users who complete a large number of classifications following every data release, increasing the disparity in user contributions (the Gini coefficient).

\subsection{User Weighting}
\label{subsec:weighting} 
User weights are calculated for each individual volunteer in order to identify users who are more sensitive to detecting transit-like signals and those who are more likely to mark false positives. The weighting scheme is based on the weighting scheme described by \cite{schwamb12}.

User weights are calculated independently for each observation sector, using the simulated light curves shown alongside the data from that sector. All users start off with a weighting of one, which is then increased or decreased when a simulated transit event is correctly or incorrectly identified, respectively. 

Simulated transits are deemed correctly identified, or `True', if the mid-point of a user's marking falls within the width of the simulated transit events. If none of the user's markings fall within this range, the simulated transit is deemed not identified, or `False'. If more than one of a user's markings coincide with the same simulated signal, it is only counted as being correct once, such that the total number of `True' markings cannot exceed the number of injected signals. For each classification, we record the number of  `Extra' markings, which is the total number of markings made by the user minus the number of correctly identified simulated transits. 

Each simulated light curve, identified by superscript $i$ (where $i=1$, \ldots, $N$) was seen by $K^{(i)}$ users (the mean value of $K^{(i)}$
was 10), and contained $T^{(i)}$ simulated transits (where $T^{(i)}$ depends on the period of the simulated transit signal and the duration of the light curve). For a specific light curve $i$, each user who saw the light curve is identified by a subscript $k$ (where $k=1$, \ldots, $K^{(i)}$) and each injected transit by a subscript $t$ (where $t=1$, \ldots, $T^{(i)}$). 

In order to distinguish between users who are able to identify obvious transits and those who are also able to find those that are more difficult to see, we start by defining a `recoverability' $r^{(i)}_t$ for each injected transit $t$ in each light curve. This is defined empirically, as the number of users who identified the transit correctly divided by $K^{(i)}$ (the total number of users who saw the light curve in question).

Next, we quantify the performance of each user on each light curve as follows (this performance is analogous to the `seed' defined in \citealt{schwamb12}, but we define it slightly differently):
\begin{equation}
  p^{(i)}_{k} = C_{\rm E} ~ \frac{E^{(i)}_{k}}{\langle E^{(i)} \rangle} + \sum_{t=1}^{T^{(i)}} \begin{cases}
    C_{\rm T} ~ \left[ r^{(i)}_t \right]^{-1}, & \text{if $m^{(i)}_{t,k} = $`True'}\\
    C_{\rm F} ~ r^{(i)}_t, & \text{if $m^{(i)}_{t,k} = $`False'},
  \end{cases}
\end{equation}
where $m^{(i)}_{t,k}$ is the identification of transit $t$ by user $k$ in light curve $i$, which is either `True' or `False'; $E^{(i)}_{k}$ is the number of `Extra' markings made by user $k$ for light curve $i$, and  $\langle E^{(i)} \rangle$ is the mean number of `Extra' markings made by all users who saw subject $i$. The parameters $C_{\rm E}$, $C_{\rm T}$ and $C_{\rm F}$ control the impact of the `Extra', `True' and `False' markings on the overall user weightings, and are optimized empirically as discussed below in Section~\ref{subsec:optimizesearch}. 

Following \citealt{schwamb12}, we then assign a global `weight' $w_k$ to each user $k$, which is defined as:
\begin{equation}
\begin{split}
	w_k = I \times (1 + \log_{10} N_k)^{\nicefrac{\sum_i p^{(i)}_k}{N_k}}
\label{equ:weight}
\end{split}
\end{equation}
where $I$ is an empirical normalization factor, such that the distribution of user weights remains centred on one, $N_k$ is the total number of simulated transit events that user $k$ assessed, and the sum over $i$ concerns only the light curves that user $k$ saw. 
We limit the user weights to the range 0.05--3 \emph{a posteriori}.


We experimented with a number of alternative ways to define the user weights, including the simpler $w_k=\nicefrac{\sum_i p^{(i)}_k}{N_k}$, but Eqn.~\ref{equ:weight} was found to give the best results (see Section~\ref{sec:recovery_efficiency} for how this was evaluated).

\subsection{Systematic Removal}
\label{subsec:sysrem} 
\begin{figure}
    \centering
    \includegraphics[width=0.46\textwidth]{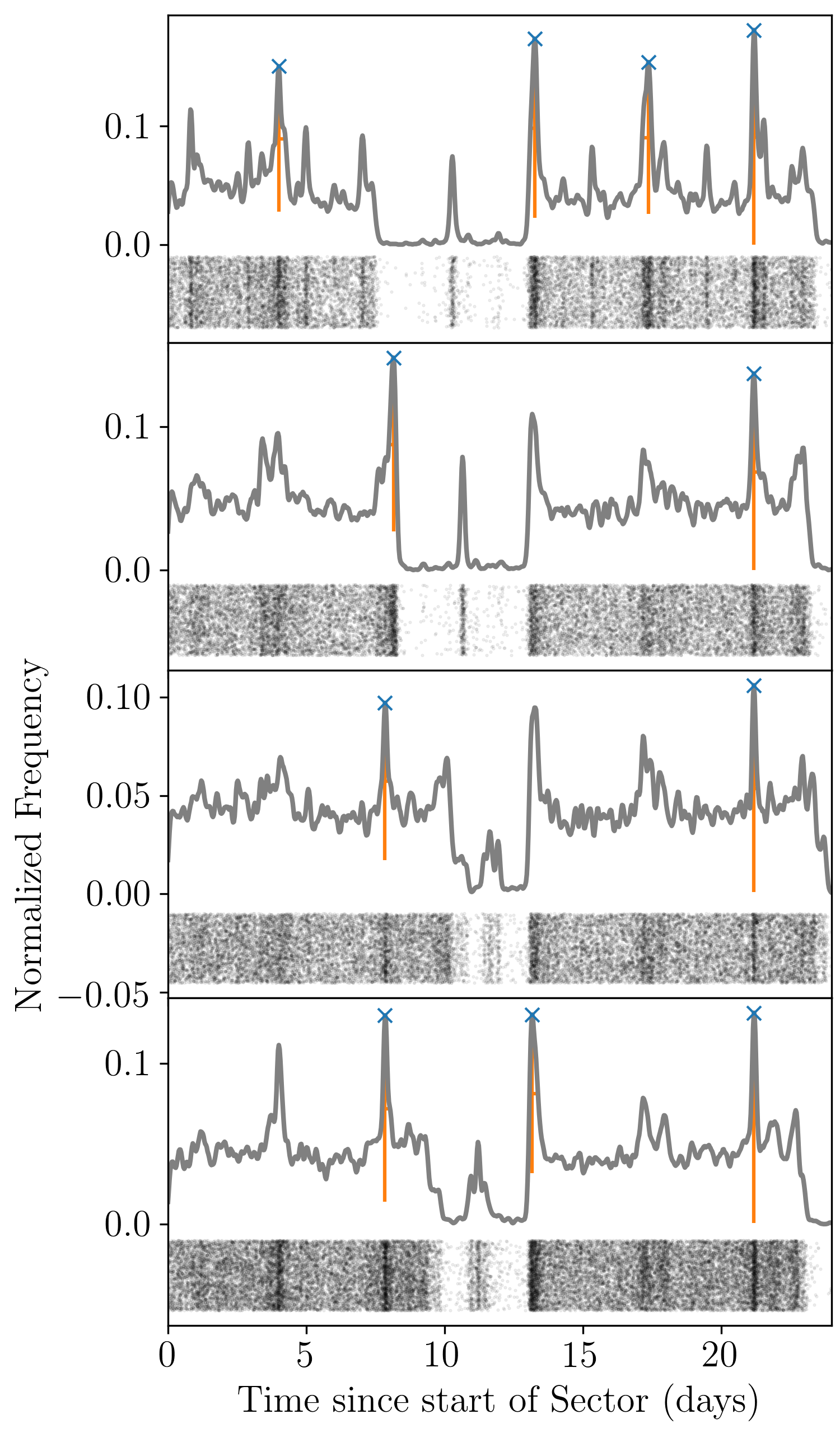}
    \caption{
    Kernel density estimation of the user-markings made for Sector 17, for targets observed with TESS's observational Cameras 1 (top panel) to 4 (bottom panel). The orange vertical lines the indicate prominences that are at least four times greater than the standard deviation of the distribution. The black points underneath the figures show the mid-points of all of the volunteer-markings, where darker regions represent a higher density of markings.}
    \label{fig:sys_rem}
\end{figure}

Systematic effects, for example caused by the spacecraft or background events, can result in spurious signals that affect a large subset of the data, resulting in an excess in markings of transit-like events at certain times within an observation sector. As the four \tess\ cameras can yield unique systematic effects, the times of systematics were identified uniquely for each camera. The times were identified using a Kernel Density Estimation \citep[KDE;][]{rosenblatt1956} with a cosine kernel and a bandwidth of 0.1 days, applied across all of the markings from that sector for each camera. Fig.~\ref{fig:sys_rem} shows the KDE of all marked transit-events made during Sector 17 for TESS's cameras 1 (top panel) to 4 (bottom panel). The isolated spikes, or prominences, in the number of marked events, such as at T = 21-22 days in the bottom panel, are assumed to be caused by systematic effects that affect multiple light curves. Prominences are considered significant if they exceed a factor four times the standard deviation of the kernel output, which was empirically determined to be the highest cut-off to not miss clearly visible systematics. All user-markings within the full width at half maximum of these peaks are omitted from all further analysis. \textcolor{black}{The KDE profiles for each Sector are provided as electronic supplementary material.}

\subsection{Density Based Clustering}
\label{subsec:DBscan} 

The times and likelihoods of transit-like events are determined by combining all of the classifications made for each subject and identifying times where multiple volunteers identified a signal. We do this using an unsupervised machine learning method, known as DBSCAN \citep[][Density-Based Spatial Clustering of Applications with Noise]{ester1996DB}. DBSCAN is a non-parametric density based clustering algorithm that helps to distinguish between dense clusters of data and sparse noise. For a data point to belong to a cluster it must be closer than a given distance ($\epsilon$) to at least a set minimum number of other points (minPoints). 

In our case, the data points are one-dimensional arrays of times of transits events, as identified by the volunteers, and clusters are times where multiple volunteers identified the same event. For each cluster a `transit score' ($s_i$) is determined, which is the sum of the user weights of the volunteers who contribute to the given cluster divided by the sum of the user weights of volunteers who saw that light curve. These transit scores are used to rank subjects from most to least likely to contain a transit-like event. Subjects which contain multiple successful clusters with different scores are ranked by the highest transit score. 

\subsection{Optimizing the search}
\label{subsec:optimizesearch}

The methodology described in Sections~\ref{subsec:weighting} to \ref{subsec:DBscan} has five free parameters: the number of markings required to constitute a cluster ($minPoints$), the maximum separation of markings required for members of a cluster ($\epsilon$), and $C_{\rm E}$, $C_{\rm T}$ and $C_{\rm F}$ used in the weighting scheme. The values of these parameters were optimized via a grid search, where $C_{\rm E}$ and $C_{\rm F}$ ranged from -5 to 0, $C_{\rm T}$ ranged from 0 to 20, and $minPoints$ ranged from 1 to 8, all in steps of 1. ($\epsilon$) ranged from 0.5 to 1.5 in steps of 0.5. This grid search was carried out on 4 sectors, two from UI1 and two from UI2, for various variations of Equation~\ref{equ:weight}. 

The success of each combination of parameters was assessed by the fractions of TOIs and TCEs that were recovered within the top highest ranked 500 candidates, as discussed in more detail Section~\ref{sec:recovery_efficiency}. We found the most successful combination of parameters to be $minPoints$ = 4 markings, $\epsilon$, = 1 day, $C_{\rm T}$ = 3, $C_{\rm F}$= -2 and $C_{\rm E}$ = -2.

\subsection{MAST deliverables}
\label{subsec:deliverables}

The analysis described above is carried out both in real-time as classifications are made, as well as offline after all of the light curves of a given sector have been classified. When the real-time analysis identifies a successful DB cluster (i.e. when at least four citizen scientists identified a transit within a day of the \tess\ data of one another), the potential candidate is automatically uploaded to the open access Planet Hunters Analysis Database (PHAD) \footnote{\url{https://mast.stsci.edu/phad/}} hosted by the Mikulski Archive for Space Telescopes (MAST) \footnote{\url{https://archive.stsci.edu/}}. While PHAD does not list every single classification made on PHT, it does display all transit candidates which had significant consensus amongst the volunteers who saw that light curve, along with the user-weight-weighted transit scores. This analysis does not apply the systematics removal described in Section~\ref{subsec:sysrem}. The aim of PHAD is to provide an open source database of potential planet candidates identified by PHT, and to credit the volunteers who identified said targets. 

The offline analysis is carried out following the complete classifications of all of the data from a given \tess\ sector. The combination of all of the classifications allows us to identify and remove times of systematics and calculate better calibrated and more representative user weights. The remainder of this paper will only discuss the results from the offline analysis.

\section{Recovery Efficiency}
\label{sec:recovery_efficiency}
\subsection{Recovery of simulated transits}

\begin{figure}
    \centering
    \includegraphics[width=0.47\textwidth]{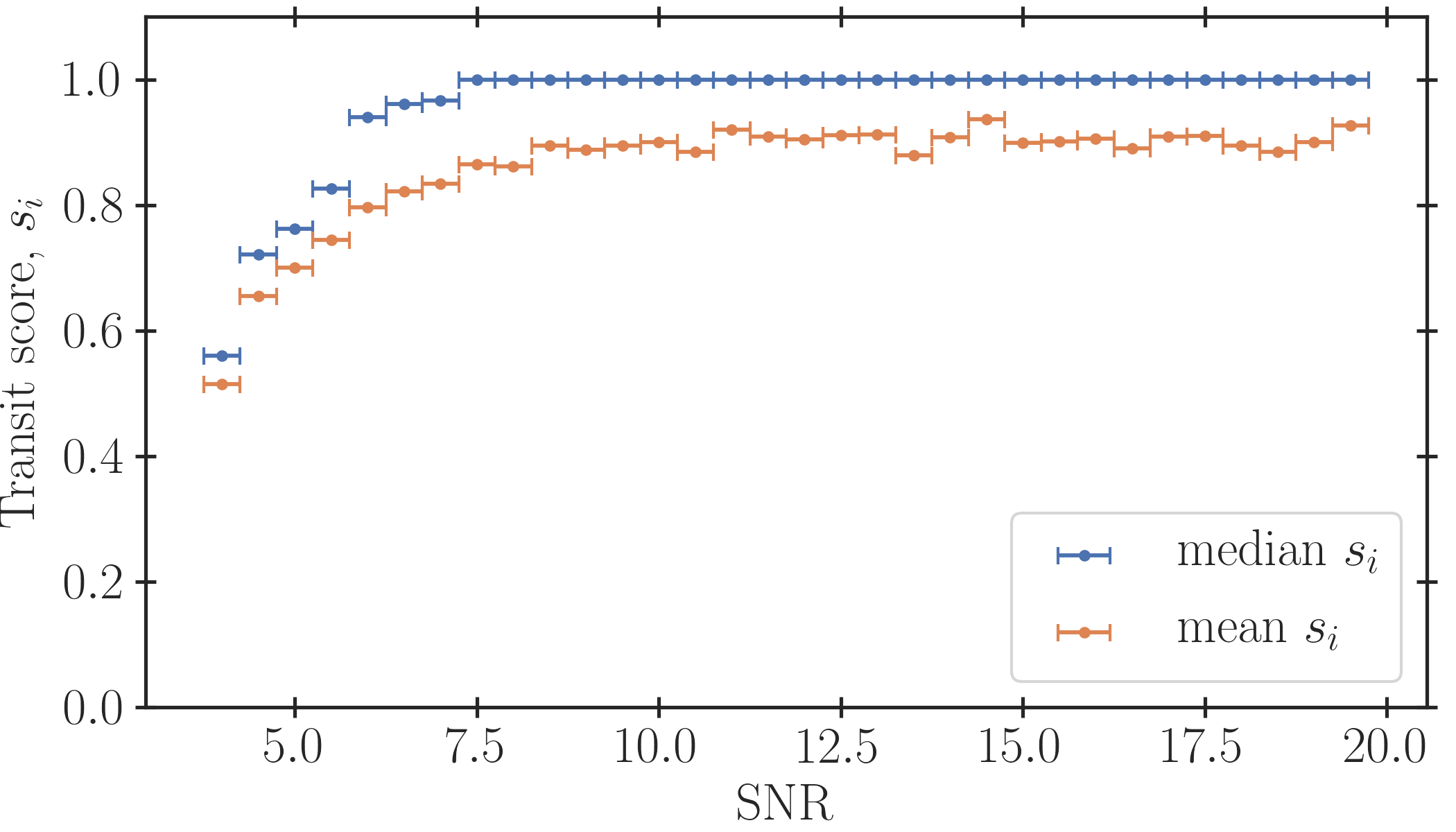}
    \caption{The median (blue) and mean (orange) transit scores for injected transits with SNR ranges between 4 and 20. The mean and median are calculated in SNR bins with a width of 0.5, as indicated by the horizontal lines around each data point. 
    }
    \label{fig:SIM_recovery}
\end{figure}

The recovery efficiency is, in part, assessed by analysing the recovery rate of the injected transit-like signals (see Section~\ref{subsec:sims}). Figure~\ref{fig:SIM_recovery} shows the median and mean transit scores (fraction of volunteers who correctly identified a given transit scaled by user weights) of the simulated transits within SNR bins ranging from 4 to 20 in steps of 0.5. Simulations with a SNR less than 4 were not shown on PHT. The figure highlights that transit signals with a SNR of 7.5 or greater are correctly identified by the vast majority of volunteers. 

\textcolor{black}{As the simulated data solely consist of real light curves with synthetically injected transit signals, we do not have any light curves, simulated or otherwise, which we can guarantee do not contain any planetary transits (real or injected). As such, this prohibits us from using simulated data to infer an analogous false-positive rate.}

\subsection{Recovery of TCEs and TOIs}
\label{subsec:TCE_TOI}
The recovery efficiency of PHT is assessed further using the planet candidates identified by the SPOC pipeline \citep{Jenkins2018}. The SPOC pipeline extracts and processes all of the 2-minute cadence \tess\ light curves prior to performing a large scale transit search. Data Validation (DV) reports, which include a range of transit diagnostic tests, are generated by the pipeline for around 1250 Threshold Crossing Events (TCEs), which were flagged as containing two or more transit-like features. Visual vetting is then performed by the \tess\ science team on these targets, and promising candidates are added to the catalog of \tess\ Objects of Interest (TOIs). Each sector yields around 80 TOIs \textcolor{black}{and a mean of 1025 TCEs.}

\begin{figure}
    \centering
    \includegraphics[width=0.45\textwidth]{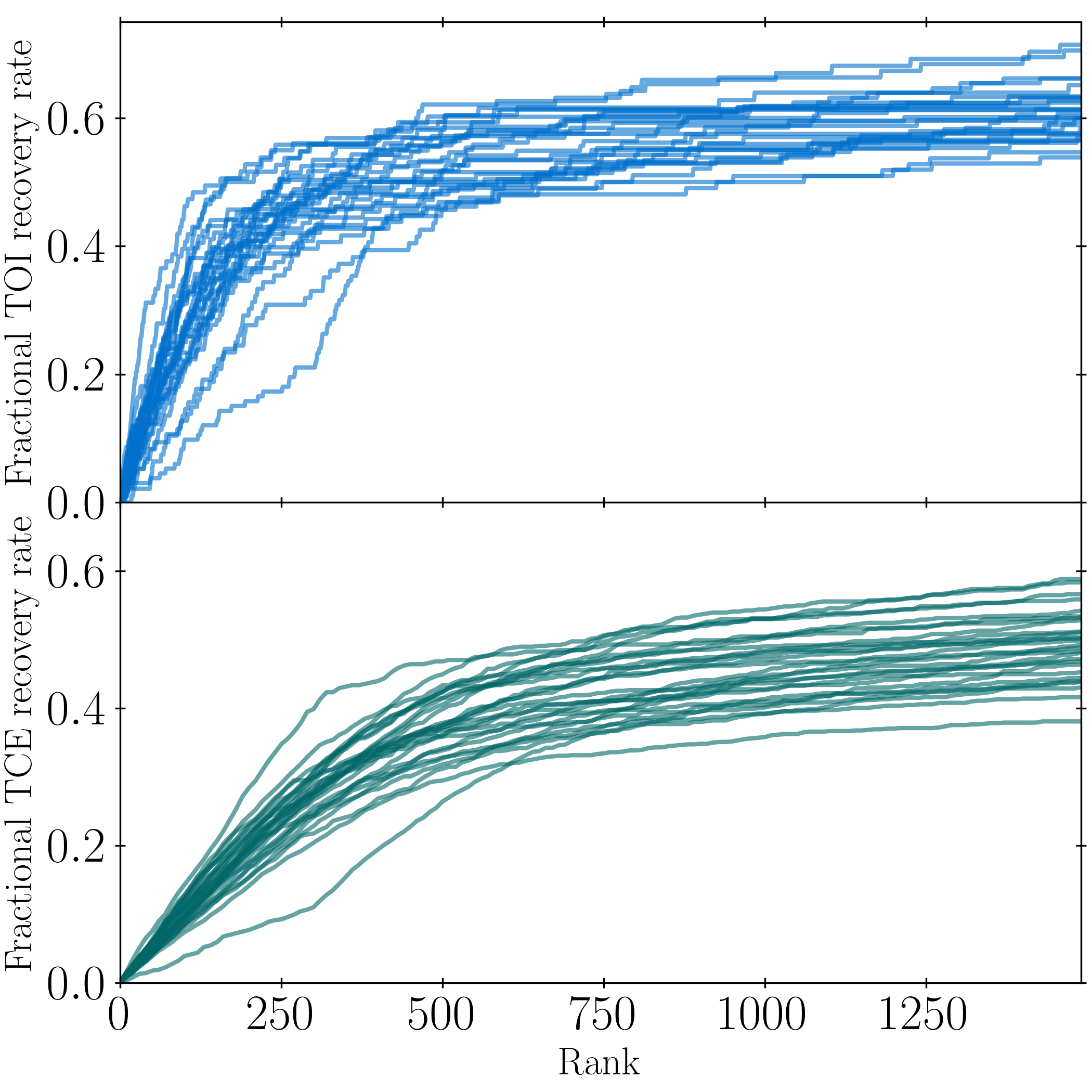}
    \caption{The fraction of recovered TOIs and TCEs (top and bottom panel respectively) with R > 2$R_{\oplus}$ as a function of the rank, for sectors 1 to 26. The lines represent the results from different observation sectors.}
    \label{fig:TCE_TOI_recovery}
\end{figure}

Fig~\ref{fig:TCE_TOI_recovery} shows the fraction of TOIs and TCEs (top and bottom panel respectively) that we recover with PHT as a function of the rank, where a higher rank corresponds to a lower transit score, for Sectors 1 through 26. TOIs and TCEs with R < 2 $R_{\oplus}$ are not included in this analysis, as the initial PH showed that human vetting alone is unable to reliably recover planets smaller than 2 $R_{\oplus}$ \citep{schwamb12}. Planets smaller than 2 $R_{\oplus}$ are, therefore, not the main focus of our search.

Fig~\ref{fig:TCE_TOI_recovery} shows a steep increase in the fractional TOI recovery rate up to a rank of $\sim$ 500. Within the 500 highest ranked PHT candidates for a given sector, we are able to recover between 46 and 62 \% (mean of 53 \%) of all of the TOIs (R > 2 $R_{\oplus}$), a median 90 \% of the TOIs where the SNR of the transit events are greater than 7.5 and median 88 \% of TOIs where the SNR of the transit events are greater than 5.

\begin{figure}
    \centering
    \includegraphics[width=0.49\textwidth]{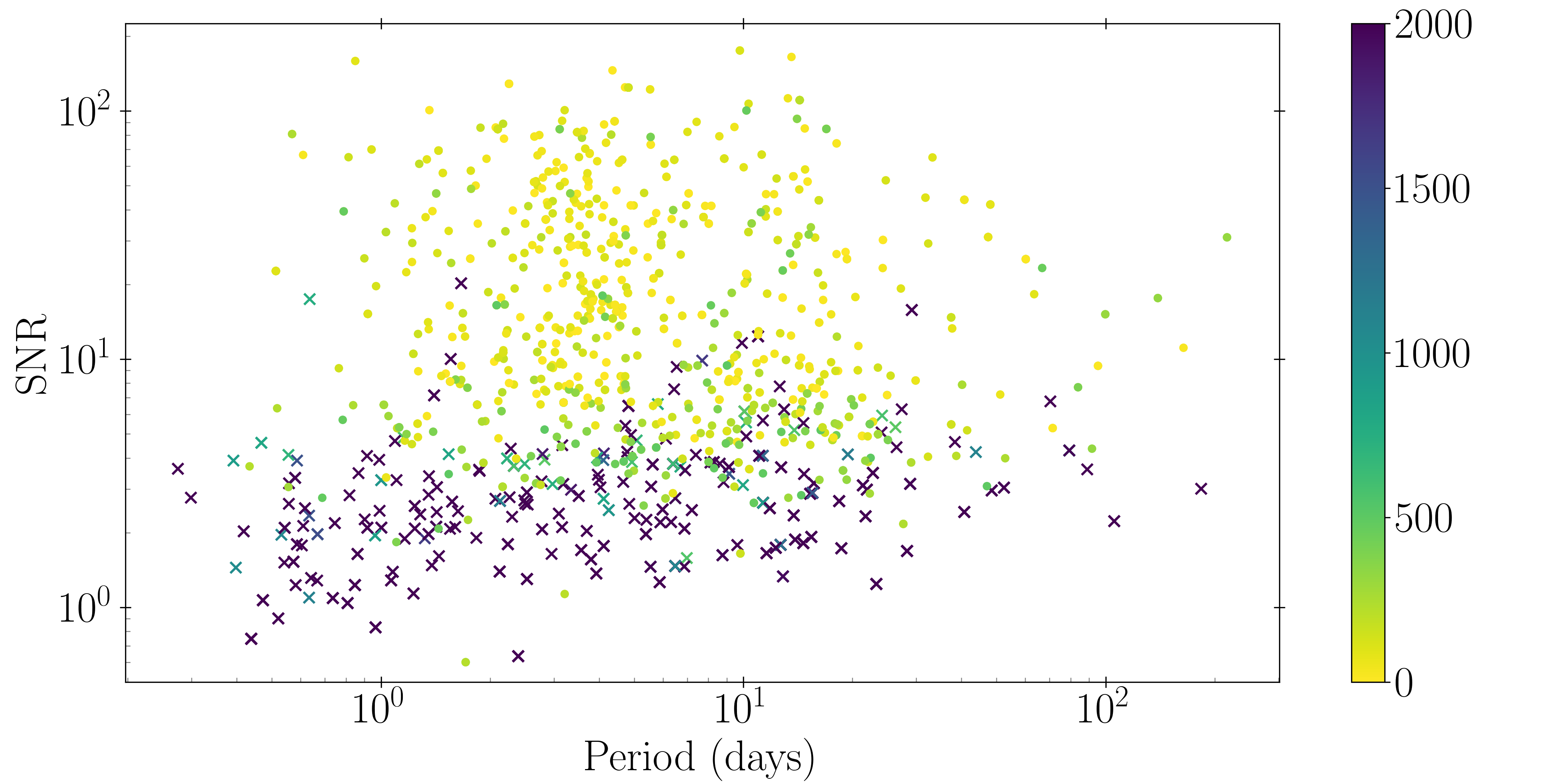}
    \caption{The SNR vs orbital period of TOIs with R > 2$R_{\oplus}$. The colour represents their rank within the sector, as determined by the weighted DB clustering algorithm. Circles indicate that they were identified at a rank < 500, while crosses indicate that they were not within the top 500 highest ranked candidates of a given sector.
    }
    \label{fig:TOI_properties}
\end{figure}

The relation between planet recovery rate and the SNR of the transit events is further highlighted in Figure~\ref{fig:TOI_properties}, which shows the SNR vs the orbital period of the recovered TOIs. The colour of the markers indicate the TOI's rank within a given sector, with the lighter colours representing a lower rank. The circles and crosses represent candidates at a rank lower and higher than 500, respectively. The figure shows that transit events with a SNR less than 3.5 are missed by the majority of volunteers, whereas events with a SNR greater than 5 are mostly recovered within the top 500 highest ranked candidates. 

The steep increase in the fractional TOI recovery rate at lower ranks, as shown in figure~\ref{fig:TCE_TOI_recovery}, is therefore due to the detection of the high SNR candidates that are identified by most, if not all, of the PHT volunteers who classified those targets. At a rank of around 500, the SNR of the TOIs tends towards the limit of what human vetting can detect and thus the identification of TOIs beyond a rank of 500 is more sporadic.

The fractional TCE recovery rate (bottom panel of Figure~\ref{fig:TCE_TOI_recovery}) is systematically lower than that of the TOIs. There are qualitative reasons as to why humans might not identify a TCE as opposed to a TOI, including that TCEs may be caused by artefacts or periodic stellar signals that the SPOC pipeline identified as a potential transit but that the human eye would either miss or be able to rule out as systematic effect. This leads to a lower recovery fraction of TCEs comparatively, an effect that is further amplified by the much larger number of TCEs.

\begin{figure*}
    \centering
    \includegraphics[width=0.9\textwidth]{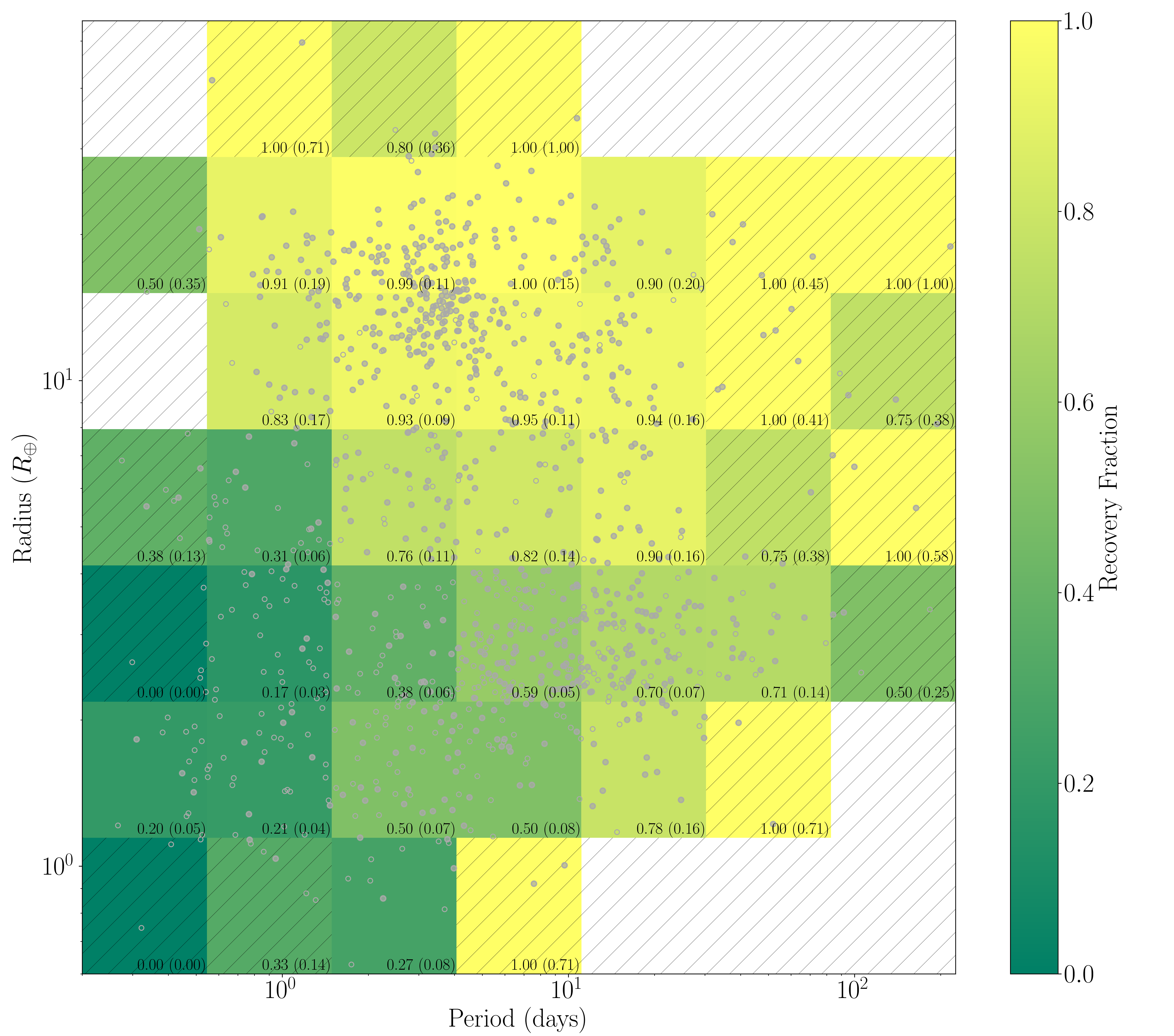}
    \caption{TOI recovery rate as a function of planet radius and orbital period. A TOI is considered recovered if it is amongst the top 500 highest ranked candidates within a given sector. The logarithmically spaced grid ranges from 0.2 to 225 d and 0.6 to 55 $R{_\oplus}$ for the orbital period and planet radius, respectively. The fraction of TOIs recovered using PHT is computed for each cell and represented by the colour the grid. Cells with less than 10 TOIs are considered incomplete for statistical analysis and are shown by the hatched lines. White cells contain no TOIs. The annotations for each cell indicate the number of recovered TOIs followed by the Poisson uncertainty in brackets. The filled in and empty grey circles indicated the recovered and not-recovered TOIs, respectively.}
    \label{fig:recovery_rank500_radius_period}
\end{figure*}

The detection efficiency of PHT is estimated using the fractional recovery rate of TOIs for a range of radius and period bins, as shown in Figure~\ref{fig:recovery_rank500_radius_period}. A TOI is considered to be recovered if its detection rank is less than 500 within the given sector. Out of the total 1913 TOIs, to date, \textcolor{black}{PHT recovered 715 TOIs among the highest ranked candidates across the 26 sectors. This corresponds to a mean of 12.7~\% of the top 500 ranked candidates per sector being TOIs. In comparison, the primary \tess\ team on average visually vets 1025 TCEs per sector, out of which a mean of 17.3~\% are promoted to TOI status.} We find that, independent of the orbital period, PHT is over 85~\% complete in the recovery of TOIs with radii equal to or greater than 4 $R_{\oplus}$. This agrees with the findings from the initial Planet Hunters project \citep{schwamb12}. The detection efficiency decreases to 51~\% for 3 - 4 $R_{\oplus}$ TOIs, 49~\% for 2 - 3 $R_{\oplus}$ TOIs and to less than 40~\% for TOIs with radii less than 2 $R_{\oplus}$. Fig~\ref{fig:recovery_rank500_radius_period} shows that the orbital period does not have a strong effect on the detection efficiency for periods greater than $\sim$~1~day, which highlights that human vetting efficiency is independent of the number of transits present within a light curve. For periods shorter than around 1~day, the detection efficiency decreases even for larger planets, due to the high frequency of events seen in the light curve. For these light curves, many volunteers will only mark a subset of the transits, which may not overlap with the subset marked by other volunteers. Due to the methodology used to identify and rank the candidates, as described in Section~\ref{sec:method}, this will actively disfavour the recovery of very short period planets. Although this obviously introduces biases in the detectability of very short period signals, the major detection pipelines are specifically designed to identify these types of planets and thus this does not present a serious detriment to our main science goal of finding planets that were \textcolor{black}{intentionally ignored or missed} by the main automated pipelines.

\begin{figure}
    \centering
    \includegraphics[width=0.45\textwidth]{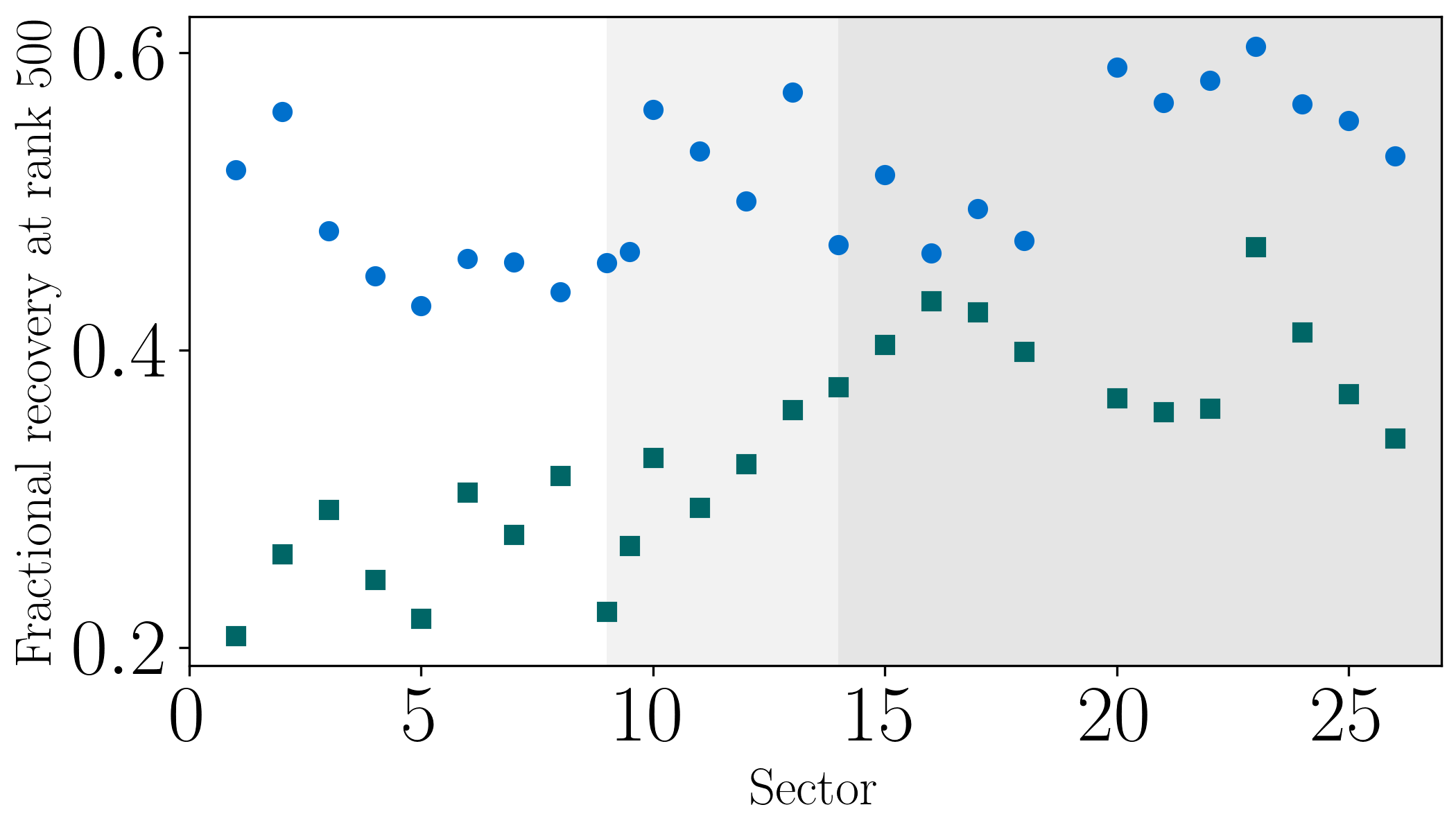}
    \caption{The fractional recovery rate of the TOIs (blue circles) and TCEs (teal squares) at a rank of 500 for each sector. Sector 1-9 (white background) represent southern hemisphere sectors classified with UI1, sectors 9-14 (light grey background) show the southern hemisphere sectors classified with UI2, and sectors 14-24 (dark grey background) show the northern hemisphere sectors classified with US2.}
    \label{fig:recovery_rank500}
\end{figure}

\begin{figure}
    \centering
    \includegraphics[width=0.50\textwidth]{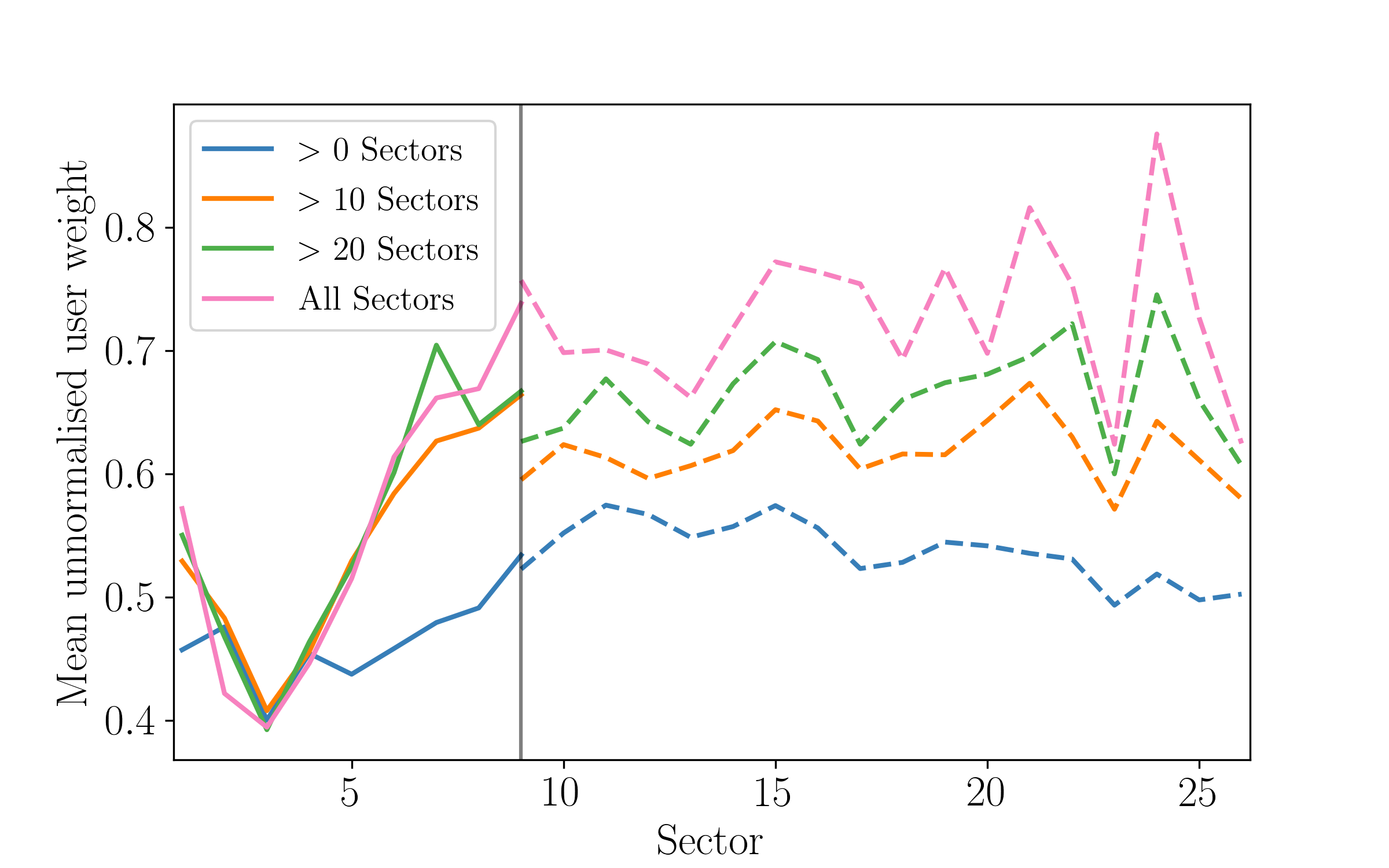}
    \caption{Mean user weights per sector. The solid lines show the user weights for the old user interface and the dashed line for the new interface, separated by the black line (Sector 9). The different coloured lines show the mean user weights calculated considering user who participated in any number of sectors (blue), more than 10 sectors (orange), more than 20 sectors (green) and all of the sectors observed during the nominal \tess\ mission (pink).}
    \label{fig:user_weights}
\end{figure}

Finally, we assessed whether the detection efficiency varies across different sectors by assessing the fraction of recovered TOIs and TCEs within the highest ranked 500 candidates. We found the recovery of TOIs within the top 500 highest ranked candidates to remain relatively constant across all sectors, while the fraction of recovered TCEs in the top 500 highest ranked candidates  increases in later sectors, as shown in Figure~\ref{fig:recovery_rank500}). After applying a Spearman's rank test we find a positive correlation of 0.86 (pvalue = 5.9 $\times$ $10^{-8}$) and 0.57 (pvalue = 0.003) between the observation sector and TCE and TOI recovery rates, respectively. These correlations suggest that the ability of users to detect transit-like events improves as they classify more subjects. The improvement of volunteers over time can also be seen in Fig~\ref{fig:user_weights}, which shows the mean (unnormalized) user weight per sector for volunteers who completed one or more classifications in at least one sector (blue), more than 10 sectors (orange), more than 20 sectors (green) and all of the sectors 26 sectors from the nominal \tess\ mission (pink). The figure highlights an overall improvement in the mean user weight in later sectors, as well as a positive correlation between the overall increase in user weight and the number of sectors that volunteers have participated in.

\section{Candidate vetting}
\label{sec:vetting}

For each observation sector the subjects are ranked according to their transit scores, and the 500 highest ranked targets (excluding TOIs) visually vetted by the PHT science team in order to identify potential candidates and rule out false positives. A vetting cut-off rank of 500 was chosen as we found this to maximise the number of found candidates while minimising the number of likely false positives. In the initial round of vetting, which is completed via a separate Zooniverse classification interface that is only accessible to the core science team, a minimum of three members of the team sort the highest ranked targets into either `keep for further analysis', `eclipsing binary' or `discard'. The sorting is based on the inspection of the full \tess\ light curve of the target, with the times of the satellite momentum dumps indicated. Additionally, around the time of each likely transit event (i.e. time of successful DB clusters) we inspect the background flux and the x and y centroid positions. Stellar parameters are provided for each candidate, subject to availability, alongside links to the SPOC Data Validation (DV) reports for candidates that had been flagged as TCEs but were never promoted to TOIs status.

Candidates where at least two of the reviewers indicated that the signal is consistent with a planetary transit are kept for further analysis. \textcolor{black}{This constitute a $\sim$~5~\% retention rate of the 500 highest ranked candidates per sector between the initial citizen science classification stage and the PHT science team vetting stage. Considering that the known planets and TOIs are not included at this stage of vetting, it is not surprising that our retention rate is lower that the true-positive rates of TCEs (see Section~\ref{subsec:TCE_TOI}). Furthermore, this false-positive rate is consistent with the the findings of the initial Planet Hunters project \citep{schwamb12}.}

The rest of the 500 candidates were grouped into $\sim$~37~\% `eclipsing binary' and $\sim$~58~\% `discard'. The most common reasons for discarding light curves are due to events caused by momentum dumps and due to background events, such as background eclipsing binaries, that mimic transit-like signals in the light curve. The targets identified as eclipsing binaries are analysed further by the \tess\ Eclipsing Binaries Working Group (Prsa et al, in prep).

For the second round of candidate vetting we generate our own data validation reports for all candidates classified as `keep for further analysis'. The reports are generated using the open source software {\sc latte} \citep[Lightcurve Analysis Tool for Transiting Exoplanets;][]{LATTE2020}, which includes a range of standard diagnostic plots that are specifically designed to help identify transit-like signals and weed out astrophysical false positives in \tess\ data. In brief the diagnostics consist of:

\textbf{Momentum Dumps}. The times of the \tess\ reaction wheel momentum dumps that can result in instrumental effects that mimic astrophysical signals.

\textbf{Background Flux}. The background flux to help identify trends caused by background events such as asteroids or fireflies \citep{vanderspek2018tess} passing through the field of view.

\textbf{x and y centroid positions}. The CCD column and row local position of the target's flux-weighted centroid, and the CCD column and row motion which considers differential velocity aberration (DVA), pointing drift, and thermal effects. This can help identify signals caused by systematics due to the satellite. 
\textbf{Aperture size test}. The target light curve around the time of the transit-like event extracted using two apertures of different sizes. This can help identify signals resulting from background eclipsing binaries.
 
\textbf{Pixel-level centroid analysis}. A comparison between the average in-transit and average out-of-transit flux, as well as the difference between them. This can help identify signals resulting from background eclipsing binaries.

\textbf{Nearby companion stars}. The location of nearby stars brighter than V-band magnitude 15 as queried from the Gaia Data Release 2 catalog \citep{gaia2018gaia} and the DSS2 red field of view around the target star in order to identify nearby contaminating sources. 

\textbf{Nearest neighbour light curves}. Normalized flux light curves of the five short-cadence \tess\ stars with the smallest projected distances to the target star, used to identify alternative sources of the signal or systematic effects that affect multiple target stars. 

\textbf{Pixel level light curves}. Individual light curves extracted for each pixel around the target. Used to identify signals resulting from background eclipsing binaries, background events and systematics.

\textbf{Box-Least-Squares fit}. Results from two consecutive BLS searches, where the identified signals from the initial search are removed prior to the second BLS search.

The {\sc latte} validation reports are assessed by the PHT science team in order to identify planetary candidates that warrant further investigation. Around 10~\% of the targets assessed at this stage of vetting are kept for further investigation, resulting in $\sim$~3 promising planet candidates per observation sector. The discarded candidates can be loosely categorized into (background) eclipsing binaries ($\sim$~40~\%), systematic effects ($\sim$~25~\%), background events ($\sim$~15~\%) and other (stellar signals such as spots; $\sim$~10~\%).

We use \pyaneti\ \citep{pyaneti} to infer the planetary and orbital parameters of our most promising candidates. For multi-transit candidates we fit for seven parameters per planet, time of mid-transit $T_0$, orbital period $P$, impact parameter $b$, scaled semi-major axis $a/R_\star$, scaled planet radius $r_{\rm p}/R_\star$, and two limb darkening coefficients following a \citet{Mandel2002} quadratic limb darkening model, implemented with the $q_1$ and $q_2$ parametrization suggested by \citet{Kipping2013}. Orbits were assumed to be circular.
For the mono-transit candidates, we fit the same parameters as for the multi-transit case, except for the orbital period and scaled semi-major axis which cannot be known for single transits. We follow \citet{Osborn2016} to estimate the orbital period of the mono-transit candidates assuming circular orbits.

We note that some of our candidates are V-shaped, consistent with a grazing transit configuration. For these cases, we set uniform priors between 0 and 0.15 for $r_{\rm p}/R_\star$ and between 0 and 1.15 for the impact parameter in order to avoid large radii caused by the $r_{\rm p}/R_\star - b$ degeneracy. Thus, the $r_{\rm p}/R_\star$ for these candidates should not be trusted. A full characterisation of these grazing transits is out of the scope of this manuscript.

Figure~\ref{fig:PHT_pyaneti} shows the \tess\ transits together with the inferred model for each candidate. Table~\ref{tab:PHT-caniddates} shows the inferred main parameters, the values and their uncertainties are given by the median and 68.3\% credible interval of the posterior distributions.



\begin{figure*}
    \centering
    \includegraphics[width=0.83\textwidth]{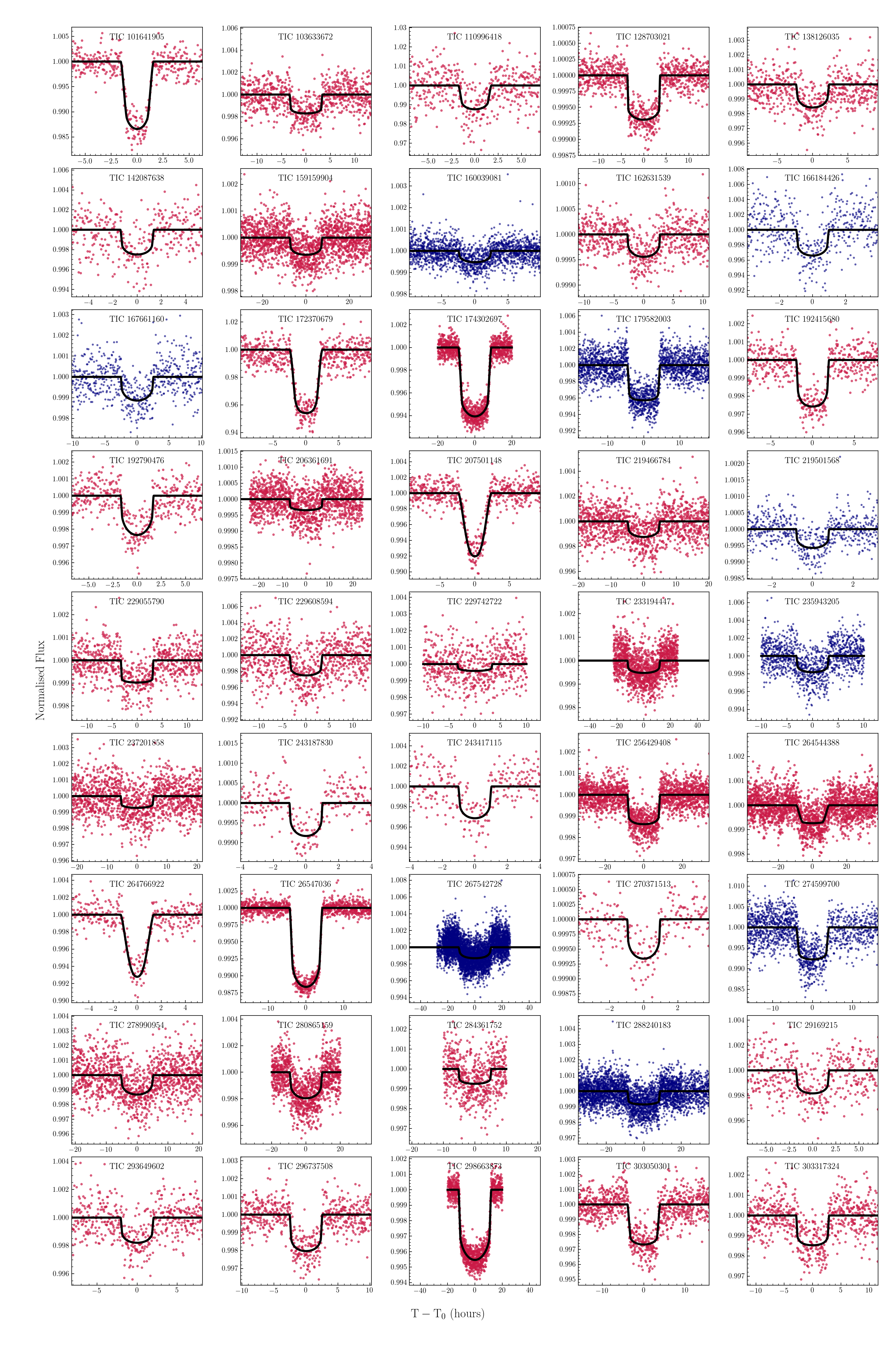}
    \caption{All of the PHT candidates modelled using \pyaneti. The parameters of the best fits are summarised in Table~\protect\ref{tab:PHT-caniddates}. The blue and magenta fits show the multi and single transit event candidates, respectively.} 
    \label{fig:PHT_pyaneti}
\end{figure*}

\begin{figure*}
    \centering
    \includegraphics[width=0.83\textwidth]{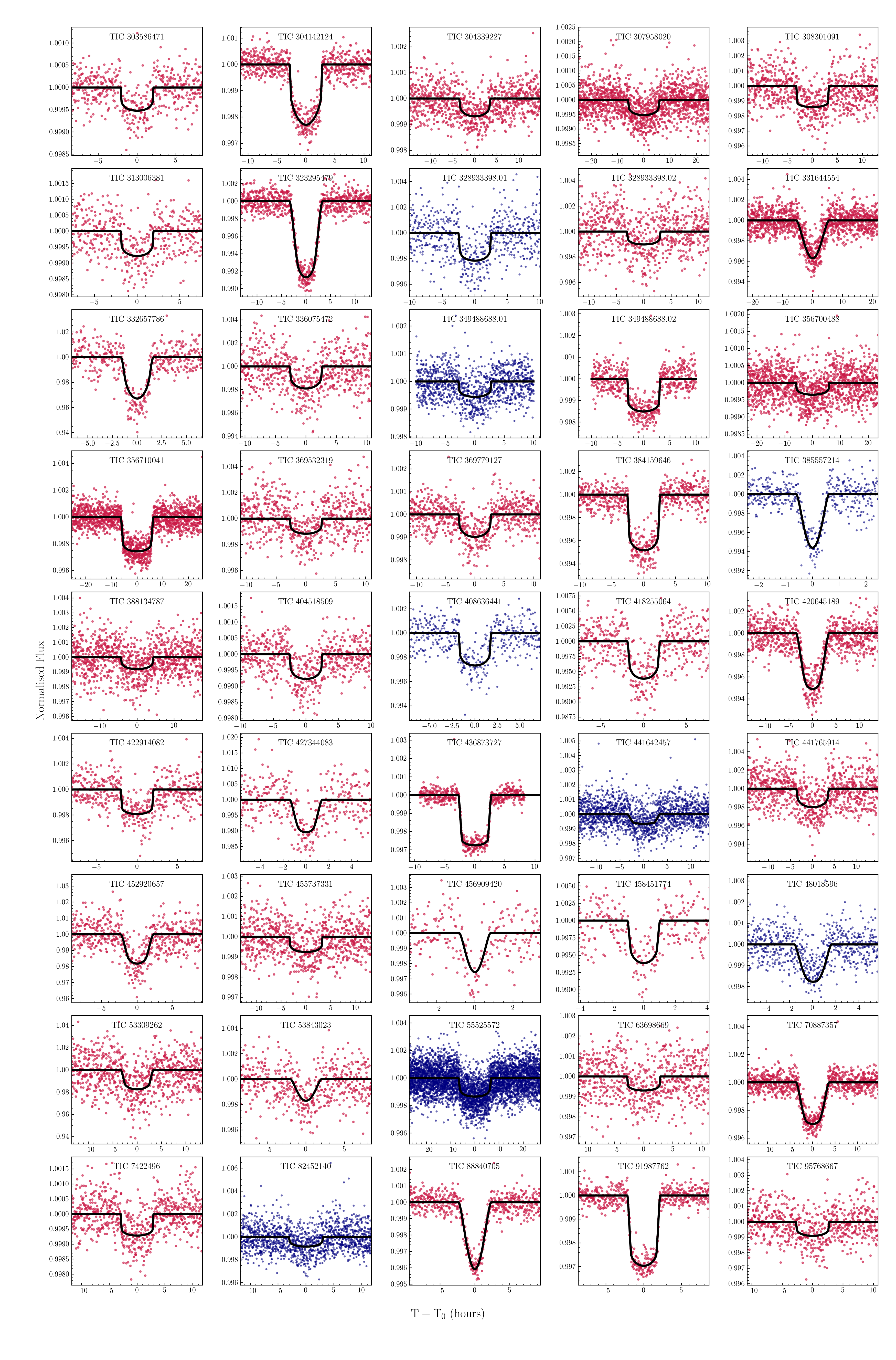}
    \addtocounter{figure}{-1}
    \caption{\textbf{PHT candidates (continued)}} 
\end{figure*}

Candidates that pass all of our rounds of vetting are uploaded to the Exoplanet Follow-up Observing Program for TESS (ExoFOP-TESS) website\footnote{\url{ https://exofop.ipac.caltech.edu/tess/index.php}} as community TOIs (cTOIs).

\section{Follow-up observations}
\label{sec:follow_up}

Many astrophysical false positive scenarios can be ruled out from the detailed examination of the \tess\ data, both from the light curves themselves and from the target pixel files. However, not all of the false positive scenarios can be ruled out from these data alone, due in part to the large \tess\ pixels (20 arcsconds). Our third stage of vetting, therefore, consists of following up the candidates with ground based observations including photometry, reconnaissance spectroscopy and speckle imaging. The results from these observations will be discussed in detail in a dedicated follow-up paper. 

\subsection{Photometry}

We make use of the LCO global network of fully robotic 0.4-m/SBIG and 1.0-m/Sinistro facilities \citep{LCO2013} to observe additional transits, where the orbital period is known, in order to refine the ephemeris and confirm that the transit events are not due to a blended eclipsing binary in the vicinity of the main target. Snapshot images are taken of single transit event candidates in order to identify nearby contaminating sources. 


\subsection{Spectroscopy}

We perform high-resolution optical spectroscopy using telescopes from across the globe in order to cover a wide range of RA and Dec:
\begin{itemize}
\item The Las Cumbres Observatory (LCO) telescopes with the Network of Robotic Echelle Spectrographs \citep[NRES,][]{LCO2013}. These fibre-fed spectrographs, mounted on 1.0-m telescopes around the globe, have a resolution of R = 53,000 and a wavelength coverage of 380 to 860 nm. 

\item The MINERVA Australis Telescope facility, located at Mount Kent Observatory in Queensland, Australia \citep{addison2019}. This facility is made up of four 0.7m CDK700 telescopes, which individually feed light via optic fibre into a KiwiSpec high-resolution (R = 80,000) stabilised spectrograph \citep{barnes2012} that covers wavelengths from 480 nm to 620 nm. 

\item The CHIRON spectrograph mounted on the SMARTS 1.5-m telescope \citep{Tokovinin2018}, located at the Cerro Tololo
Inter-American Observatory (CTIO) in Chile. The high resolution cross-dispersed echelle spectrometer is fiber-fed followed by an image slicer. It has a resolution of R = 80,000 and covers wavelengths ranging from 410 to 870 nm.

\item The SOPHIE echelle spectrograph mounted on the 1.93-m Haute-Provence Observatory (OHP), France
\citep{2008Perruchot,2009Bouchy}. The high resolution cross-dispersed stabilized echelle spectrometer is fed by two optical fibers. Observations were taken in high-resolution mode (R = 75,000) with a wavelength range of 387 to 694 nm.

\end{itemize}

Reconnaissance spectroscopy with these instruments allow us to extract stellar parameters, identify spectroscopic binaries, and place upper limits on the companion masses. Spectroscopic binaries and targets whose spectral type is incompatible with the initial planet hypothesis and/or precludes precision RV observations (giant or early type stars) are not followed up further. Promising targets, however, are monitored in order to constrain their period and place limits on their mass. 

\subsection{Speckle Imaging}

For our most promising candidates we perform high resolution speckle imaging using the `Alopeke instrument on the 8.1-m Frederick C. Gillett Gemini North telescope in Maunakea, Hawaii, USA, and its twin, Zorro, on the 8.1-m Gemini South telescope on Cerro Pach\'{o}n, Chile \citep{Matson2019, Howell2011}. Speckle interferometric observations provide extremely high resolution images reaching the diffraction limit of the telescope. We obtain simultaneous 562 nm and 832 nm rapid exposure (60 msec) images in succession that effectively `freeze out' atmospheric turbulence and through Fourier analysis are used to search for close companion stars at 5-8 magnitude contrast levels. This analysis, along with the reconstructed images, allow us to identify nearby companions and to quantify their light contribution to the TESS aperture and thus the transit signal.

\section{Planet candidates and Noteworthy Systems}
\label{sec:PHT_canidates}
\subsection{Planet candidate properties}

In this final part of the paper we discuss the 90 PHT candidates around 88 host stars that passed the initial two stages of vetting and that were uploaded to ExoFOP as cTOIs. At the time of discovery none of these candidates were TOIs. The properties of all of the PHT candidates are summarised in Table~\ref{tab:PHT-caniddates}. Candidates that have been promoted to TOI status since their PHT discovery are highlighted with an asterisk following the TIC ID, and candidates that have been shown to be false positives, based on the ground-based follow-up observations, are marked with a dagger symbol ($\dagger$). The majority (81\%) of PHT candidates are single transit events, indicated by an `s' following the orbital period presented in the table. \textcolor{black}{18 of the PHT candidates were flagged as TCEs by the \tess\ pipeline, but not initially promoted to TOI status. The most common reasons for this was that the pipeline identified a single-transit event as well as times of systematics (often caused by momentum dumps), due to its two-transit minimum detection threshold. This resulted in the candidate being discarded on the basis of it not passing the `odd-even' transit depth test. Out of the 18 TCEs, 14 have become TOI's since the PHT discovery. More detail on the TCE candidates can be found in Appendix~\ref{appendixA}.} 

All planet parameters (columns 2 to 8) are derived from the \pyaneti\ modelling as described in Section~\ref{sec:vetting}. Finally, the table summarises the ground-based follow-up observations (see Sec~\ref{sec:follow_up}) that have been obtained to date, where the bracketed numbers following the observing instruments indicate the number of epochs. Unless otherwise noted, the follow-up observations are consistent with a planetary scenario. More in depth descriptions of individual targets for which we have additional information to complement the results in Table~\ref{tab:PHT-caniddates} can be found in Appendix~\ref{appendixA}.

\subsection{Planet candidate analysis}

\begin{figure}
    \centering
    \includegraphics[width=0.45\textwidth]{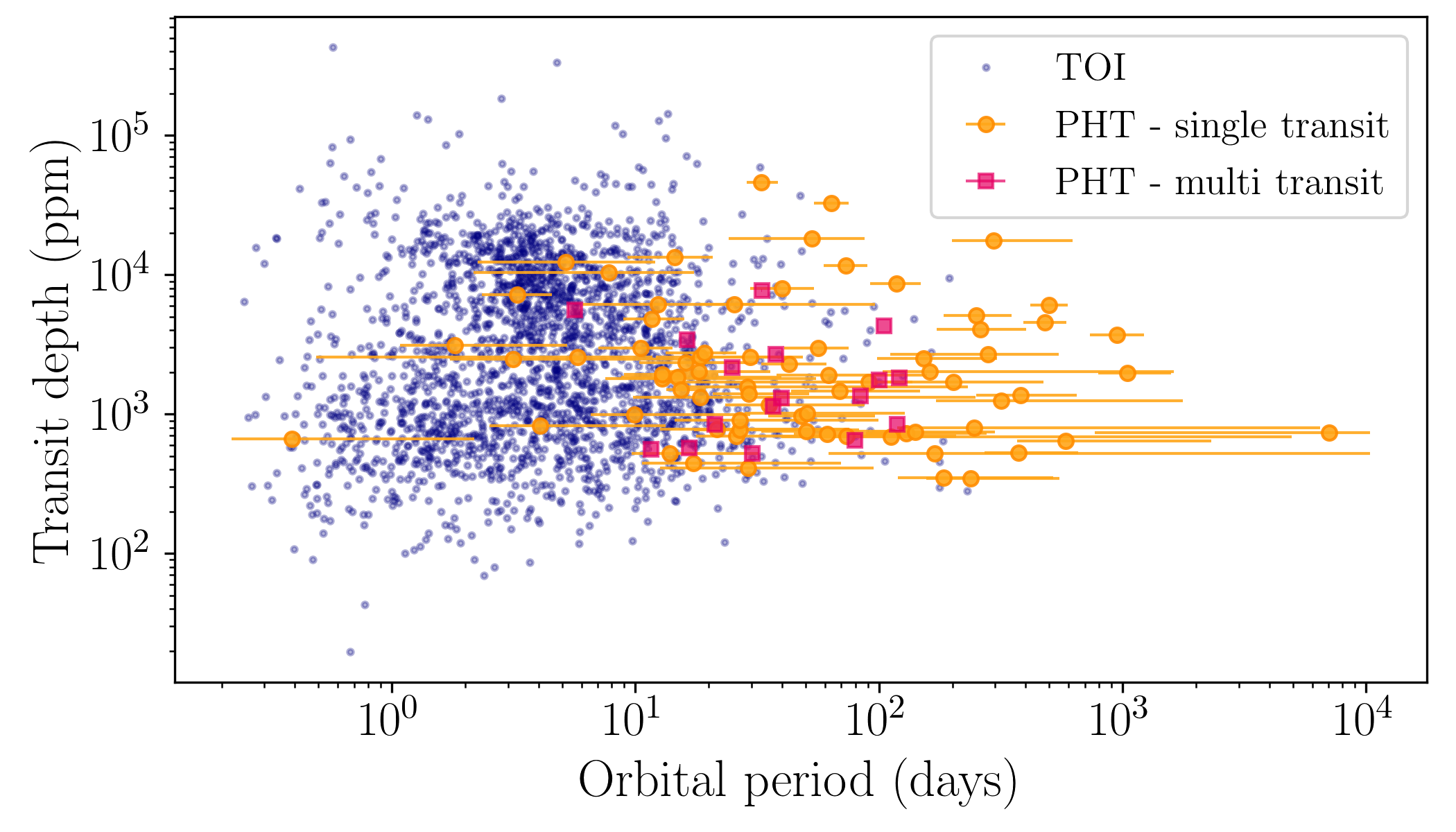}
    \caption{The properties of the PHT single transit (orange circles) and multi transit (magenta squares) candidates compared to the properties of the TOIs (blue circles). All parameters (listed in Table~\ref{fig:PHT_candidates}) were extracted using \pyaneti\ modelling.}
    \label{fig:PHT_candidates}
\end{figure}

The majority of the TOIs (87.7\%) have orbital periods shorter than 15 days due to the requirement of observing at least two transits included in all major pipelines \textcolor{black}{combined with the observing strategy of \tess}. As visual vetting does not impose these limits, the candidates outlined in this paper are helping to populate the relatively under-explored long-period region of parameter space. This is highlighted in Figure~\ref{fig:PHT_candidates}, which shows the transit depths vs the orbital periods of the PHT single transit candidates (orange circles) and the multi-transit candidates (magenta squares) compared to the TOIs (blue circles). Values of the orbital periods and transit depths were obtained via transit modelling using \pyaneti (see Section~\ref{sec:vetting}). The orbital period of single transit events are poorly constrained, which is reflected by the large errorbars in Figure~\ref{fig:PHT_candidates}. Figure~\ref{fig:PHT_candidates} also highlights that with PHT we are able to recover a similar range of transit depths as the pipeline found TOIs, as was previously shown in Figure~\ref{fig:recovery_rank500_radius_period}.

\begin{figure}
    \centering
    \includegraphics[width=0.45\textwidth]{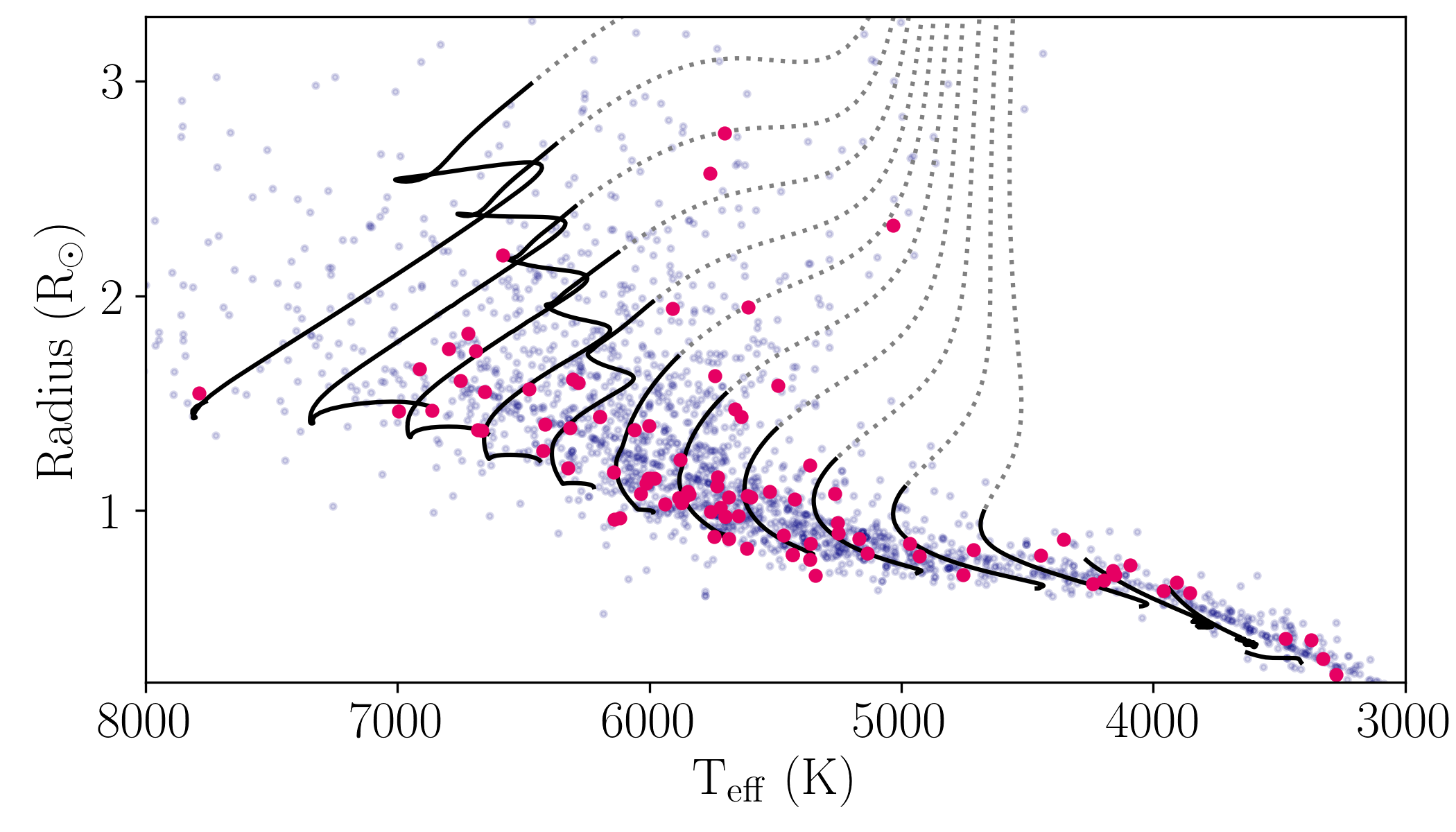}
    \caption{Stellar evolution tracks showing main sequence (solid black lines) and post-main sequence (dashed grey lines) MIST stellar evolution for stellar masses ranging from 0.3 to 1.6 $M_\odot$ in steps of 0.1 $M_\odot$. The blue dots show the TOIs and the magenta circles show the PHT candidates.} 
    \label{fig:eep}
\end{figure}

The PHT candidates were further compared to the TOIs in terms of the properties of their host stars. Figure~\ref{fig:eep} shows the effective temperature and stellar radii as taken from the TIC \citep{Stassun18}, for TOIs (blue dots) and the PHT candidates (magenta circles). The solid and dashed lines indicate the main sequence and post-main sequence MIST stellar evolutionary tracks \citep{choi2016}, respectively, for stellar masses ranging from 0.3 to 1.6 $M_\odot$ in steps of 0.1 $M_\odot$. This shows that around 10\% of the host stars are in the process of, or have recently evolved off the main sequence. The models assume solar metalicity, no stellar rotation and no additional internal mixing.

\textcolor{black}{Ground based follow-up spectroscopy has revealed that six of the PHT candidates listed in Table~\ref{tab:PHT-caniddates} are astrophysical false positives. As the follow-up campaign of the targets is still underway, the true false-positive rate of the candidates to have made it through all stages of the vetting process, as outlined in the methodology, will be be assessed in future PHT papers once the true nature of more of the candidates has been independently verified.}

\subsection{Stellar systems}
\label{subsec:PHT_stars}

In addition to the planetary candidates, citizen science allows for the identification of interesting stellar systems and astrophysical phenomena, in particular where the signals are aperiodic or small compared to the dominant stellar signal. These include light curves that exhibit multiple transit-like signals, possibly as a result of a multiple stellar system or a blend of eclipsing binaries. We have investigated all light curves that were flagged as possible multi-stellar systems via the PHT discussion boards. Similar to the planet vetting, as described in Section~\ref{sec:vetting}, we generated {\sc latte} data validation reports in order to assess the nature of the signal. Additionally, we subjected these systems to an iterative signal removal process, whereby we phase-folded the light curve on the dominant orbital period, binned the light curve into between 200-500 phase bins, created an interpolation model, and then subtracted said signal in order to evaluate the individual transit signals. The period of each signal, as listed in Table~\ref{tab:PHT-multis}, was determined by phase folding the light curve at a number of trial periods and assessing by eye the best fit period and corresponding uncertainty.

Due to the large \tess\ pixels, blends are expected to be common. We searched for blends by generating phase folded light curves for each pixel around the source of the target in order to better locate the source of each signal. Shifts in the \tess\ x and y centroid positions were also found to be good indicators of visually separated sources. Nearby sources with a magnitude difference greater than 5 mags were ruled out as possible contaminators. We consider a candidate to be a confirmed blend when the centroids are separated by more than 1 \tess\ pixel, as this corresponds to an angular separation > 21 arcseconds meaning that the systems are highly unlikely to be gravitationally bound. Systems where the signal appears to be coming from the same \tess\ pixel and that show no clear centroid shifts are considered to be candidate multiple systems. We note that blends are still possible, however, without further investigation we cannot conclusively rule these out as possible multi stellar systems. 

All of the systems are summarised in Table~\ref{tab:PHT-multis}. Out of the 26 systems, 6 are confirmed multiple systems which have either been published or are being prepared for publication; 7 are visually separated eclipsing binaries (confirmed blends); and 13 are candidate multiple system. Additional observations will be required to determine whether or not these candidate multiple systems are in fact gravitationally bound or photometric blends as a results of the large \tess\ pixels or due to a line of sight happenstance. 

\begin{landscape}
\begin{table}
\resizebox{1.31\textwidth}{!}{
\begin{tabular}{ccccccccccccc}
\textbf{TIC} & \textbf{Other} & \textbf{Epoch} & \textbf{Period}  & \textbf{$R_{pl}$/$R_{\odot}$} & \textbf{$R_{pl}$} & \textbf{Impact} & \textbf{Duration} & \textbf{$V_{mag}$} & \textbf{Photometry} & \textbf{Spectroscopy} & \textbf{Speckle} & \textbf{Comment} \\
  & \textbf{Name} & \textbf{(\textcolor{black}{BJD - 2457000})} & \textbf{(days)}  & &  $(R_{\oplus})$ & \textbf{Parameter} & \textbf{(hours)} &   &   &   &   &    \\
\hline
101641905 & TWOMASS 11412617+3441004 & $1917.26335 _{ - 0.00072 } ^ { + 0.00071 }$ & $14.52 _{ - 5.25 } ^ { + 6.21 }(s)$ & $0.1135 _{ - 0.0064 } ^ { + 0.0032 }$ & $9.76 _{ - 0.69 } ^ { + 0.65 }$ & $0.691 _{ - 0.183 } ^ { + 0.077 }$ & $3.163 _{ - 0.088 } ^ { + 0.093 }$ & 12.196 &  &  &  &  \\
103633672* & TYC 4387-00923-1 & $1850.3211 _{ - 0.00077 } ^ { + 0.00135 }$ & $90.9 _{ - 23.7 } ^ { + 46.4 }(s)$ & $0.0395 _{ - 0.0013 } ^ { + 0.0013 }$ & $3.45 _{ - 0.24 } ^ { + 0.26 }$ & $0.3 _{ - 0.21 } ^ { + 0.26 }$ & $6.7 _{ - 0.11 } ^ { + 0.12 }$ & 10.586 &  & NRES (1) &  &  \\
110996418 & TWOMASS 12344723-1019107 & $1580.6406 _{ - 0.0038 } ^ { + 0.0037 }$ & $5.18 _{ - 2.93 } ^ { + 6.86 }(s)$ & $0.1044 _{ - 0.0067 } ^ { + 0.008 }$ & $12.7 _{ - 0.99 } ^ { + 1.15 }$ & $0.44 _{ - 0.3 } ^ { + 0.3 }$ & $3.53 _{ - 0.27 } ^ { + 0.36 }$ & 13.945 &  &  &  &  \\
128703021 & HIP 71639 & $1601.8442 _{ - 0.00108 } ^ { + 0.00093 }$ & $26.0 _{ - 8.22 } ^ { + 22.35 }(s)$ & $0.0254 _{ - 0.00049 } ^ { + 0.00072 }$ & $4.44 _{ - 0.2 } ^ { + 0.23 }$ & $0.47 _{ - 0.3 } ^ { + 0.22 }$ & $7.283 _{ - 0.091 } ^ { + 0.141 }$ & 6.06 &  & NRES (2);MINERVA (34) & Gemini &  \\
138126035 & TYC 1450-00833-1 & $1954.3229 _{ - 0.0041 } ^ { + 0.0067 }$ & $28.8 _{ - 14.0 } ^ { + 203.2 }(s)$ & $0.0375 _{ - 0.0026 } ^ { + 0.0069 }$ & $4.01 _{ - 0.35 } ^ { + 0.74 }$ & $0.58 _{ - 0.38 } ^ { + 0.35 }$ & $4.65 _{ - 0.32 } ^ { + 0.85 }$ & 10.349 &  &  &  &  \\
142087638 & TYC 9189-00274-1 & $1512.1673 _{ - 0.0043 } ^ { + 0.0034 }$ & $3.14 _{ - 1.41 } ^ { + 12.04 }(s)$ & $0.0469 _{ - 0.0035 } ^ { + 0.0063 }$ & $6.05 _{ - 0.54 } ^ { + 0.89 }$ & $0.5 _{ - 0.35 } ^ { + 0.36 }$ & $2.72 _{ - 0.23 } ^ { + 0.5 }$ & 11.526 &  &  &  &  \\
159159904 & HIP 64812 & $1918.6109 _{ - 0.0067 } ^ { + 0.0091 }$ & $584.0 _{ - 215.0 } ^ { + 1724.0 }(s)$ & $0.0237 _{ - 0.0011 } ^ { + 0.0026 }$ & $3.12 _{ - 0.22 } ^ { + 0.36 }$ & $0.49 _{ - 0.34 } ^ { + 0.35 }$ & $15.11 _{ - 0.54 } ^ { + 0.7 }$ & 9.2 &  & NRES (2) &  &  \\
160039081* & HIP 78892 & $1752.9261 _{ - 0.0045 } ^ { + 0.005 }$ & $30.19918 _{ - 0.00099 } ^ { + 0.00094 }$ & $0.0211 _{ - 0.0013 } ^ { + 0.0035 }$ & $2.67 _{ - 0.21 } ^ { + 0.43 }$ & $0.52 _{ - 0.34 } ^ { + 0.36 }$ & $4.93 _{ - 0.27 } ^ { + 0.37 }$ & 8.35 & SBIG (1) & NRES (1);SOPHIE (4) & Gemini &  \\
162631539 & HIP 80264 & $1978.2794 _{ - 0.0044 } ^ { + 0.0051 }$ & $17.32 _{ - 6.66 } ^ { + 52.35 }(s)$ & $0.0195 _{ - 0.0011 } ^ { + 0.0024 }$ & $2.94 _{ - 0.24 } ^ { + 0.38 }$ & $0.48 _{ - 0.33 } ^ { + 0.36 }$ & $5.54 _{ - 0.33 } ^ { + 0.41 }$ & 7.42 &  &  &  &  \\
166184426* & TWOMASS 13442500-4020122 & $1600.4409 _{ - 0.003 } ^ { + 0.0036 }$ & $16.3325 _{ - 0.0066 } ^ { + 0.0052 }$ & $0.0545 _{ - 0.0031 } ^ { + 0.0039 }$ & $1.85 _{ - 0.12 } ^ { + 0.15 }$ & $0.41 _{ - 0.28 } ^ { + 0.31 }$ & $1.98 _{ - 0.22 } ^ { + 0.17 }$ & 12.911 &  &  &  &  \\
167661160$\dagger$ & TYC 7054-01577-1 & $1442.0703 _{ - 0.0028 } ^ { + 0.004 }$ & $36.802 _{ - 0.07 } ^ { + 0.069 }$ & $0.0307 _{ - 0.0014 } ^ { + 0.0024 }$ & $4.07 _{ - 0.32 } ^ { + 0.43 }$ & $0.37 _{ - 0.26 } ^ { + 0.33 }$ & $5.09 _{ - 0.23 } ^ { + 0.21 }$ & 9.927 &  & NRES (9);MINERVA (4) &  & EB from MINERVA observations \\
172370679* & TWOMASS 19574239+4008357 & $1711.95923 _{ - 0.00099 } ^ { + 0.001 }$ & $32.84 _{ - 4.17 } ^ { + 5.59 }(s)$ & $0.1968 _{ - 0.0032 } ^ { + 0.0022 }$ & $13.24 _{ - 0.43 } ^ { + 0.43 }$ & $0.22 _{ - 0.15 } ^ { + 0.14 }$ & $4.999 _{ - 0.097 } ^ { + 0.111 }$ & 14.88 &  &  &  & Confirmed planet \citep{canas2020}. \\
174302697* & TYC 3641-01789-1 & $1743.7267 _{ - 0.00092 } ^ { + 0.00093 }$ & $498.2 _{ - 80.0 } ^ { + 95.3 }(s)$ & $0.07622 _{ - 0.00068 } ^ { + 0.00063 }$ & $13.34 _{ - 0.57 } ^ { + 0.58 }$ & $0.642 _{ - 0.029 } ^ { + 0.024 }$ & $17.71 _{ - 0.12 } ^ { + 0.13 }$ & 9.309 & SBIG (1) &  &  &  \\
179582003 & TYC 9166-00745-1 & $1518.4688 _{ - 0.0016 } ^ { + 0.0016 }$ & $104.6137 _{ - 0.0022 } ^ { + 0.0022 }$ & $0.06324 _{ - 0.0008 } ^ { + 0.0008 }$ & $7.51 _{ - 0.35 } ^ { + 0.35 }$ & $0.21 _{ - 0.15 } ^ { + 0.19 }$ & $9.073 _{ - 0.084 } ^ { + 0.097 }$ & 10.806 &  &  &  &  \\
192415680 & TYC 2859-00682-1 & $1796.0265 _{ - 0.0012 } ^ { + 0.0013 }$ & $18.47 _{ - 6.34 } ^ { + 21.73 }(s)$ & $0.0478 _{ - 0.0017 } ^ { + 0.0027 }$ & $4.43 _{ - 0.33 } ^ { + 0.38 }$ & $0.45 _{ - 0.31 } ^ { + 0.31 }$ & $3.94 _{ - 0.1 } ^ { + 0.12 }$ & 9.838 & SBIG (1) & SOPHIE (2) &  &  \\
192790476 & TYC 7595-00649-1 & $1452.3341 _{ - 0.0014 } ^ { + 0.002 }$ & $16.09 _{ - 5.73 } ^ { + 15.49 }(s)$ & $0.0438 _{ - 0.0018 } ^ { + 0.0026 }$ & $3.24 _{ - 0.34 } ^ { + 0.37 }$ & $0.37 _{ - 0.25 } ^ { + 0.3 }$ & $3.395 _{ - 0.099 } ^ { + 0.192 }$ & 10.772 &  &  &  &  \\
206361691$\dagger$ & HIP 117250 & $1363.2224 _{ - 0.0082 } ^ { + 0.009 }$ & $237.7 _{ - 81.0 } ^ { + 314.4 }(s)$ & $0.01762 _{ - 0.00088 } ^ { + 0.00125 }$ & $2.69 _{ - 0.19 } ^ { + 0.25 }$ & $0.43 _{ - 0.28 } ^ { + 0.32 }$ & $13.91 _{ - 0.53 } ^ { + 0.52 }$ & 8.88 &  & CHIRON (2) &  & SB2 from CHIRON \\
207501148 & TYC 3881-00527-1 & $2007.7273 _{ - 0.0011 } ^ { + 0.0011 }$ & $39.9 _{ - 10.3 } ^ { + 14.3 }(s)$ & $0.0981 _{ - 0.0047 } ^ { + 0.011 }$ & $13.31 _{ - 0.95 } ^ { + 1.56 }$ & $0.9 _{ - 0.03 } ^ { + 0.039 }$ & $4.73 _{ - 0.14 } ^ { + 0.14 }$ & 10.385 &  &  &  &  \\
219466784* & TYC 4409-00437-1 & $1872.6879 _{ - 0.0097 } ^ { + 0.0108 }$ & $318.0 _{ - 147.0 } ^ { + 1448.0 }(s)$ & $0.0332 _{ - 0.0024 } ^ { + 0.0048 }$ & $3.26 _{ - 0.31 } ^ { + 0.49 }$ & $0.55 _{ - 0.39 } ^ { + 0.34 }$ & $10.06 _{ - 0.81 } ^ { + 1.12 }$ & 11.099 &  &  &  &  \\
219501568 & HIP 79876 & $1961.7879 _{ - 0.0018 } ^ { + 0.002 }$ & $16.5931 _{ - 0.0017 } ^ { + 0.0015 }$ & $0.0221 _{ - 0.0012 } ^ { + 0.0015 }$ & $4.22 _{ - 0.3 } ^ { + 0.35 }$ & $0.41 _{ - 0.28 } ^ { + 0.31 }$ & $1.615 _{ - 0.077 } ^ { + 0.093 }$ & 8.38 &  &  &  &  \\
229055790 & TYC 7492-01197-1 & $1337.866 _{ - 0.0022 } ^ { + 0.0019 }$ & $48.0 _{ - 12.8 } ^ { + 48.4 }(s)$ & $0.0304 _{ - 0.00097 } ^ { + 0.00115 }$ & $3.52 _{ - 0.2 } ^ { + 0.24 }$ & $0.37 _{ - 0.26 } ^ { + 0.32 }$ & $6.53 _{ - 0.11 } ^ { + 0.14 }$ & 9.642 &  & NRES (2) &  &  \\
229608594 & TWOMASS 18180283+7428005 & $1960.0319 _{ - 0.0037 } ^ { + 0.0045 }$ & $152.4 _{ - 54.1 } ^ { + 152.6 }(s)$ & $0.0474 _{ - 0.0023 } ^ { + 0.0024 }$ & $3.42 _{ - 0.34 } ^ { + 0.36 }$ & $0.38 _{ - 0.26 } ^ { + 0.3 }$ & $6.98 _{ - 0.23 } ^ { + 0.37 }$ & 12.302 &  &  &  &  \\
229742722* & TYC 4434-00596-1 & $1689.688 _{ - 0.025 } ^ { + 0.02 }$ & $29.0 _{ - 16.4 } ^ { + 66.3 }(s)$ & $0.019 _{ - 0.0028 } ^ { + 0.0029 }$ & $2.9 _{ - 0.44 } ^ { + 0.48 }$ & $0.44 _{ - 0.3 } ^ { + 0.33 }$ & $4.27 _{ - 0.09 } ^ { + 0.11 }$ & 10.33 &  & NRES (8);SOPHIE (4) & Gemini &  \\
233194447 & TYC 4211-00650-1 & $1770.4924 _{ - 0.0065 } ^ { + 0.0107 }$ & $373.0 _{ - 101.0 } ^ { + 284.0 }(s)$ & $0.02121 _{ - 0.00073 } ^ { + 0.001 }$ & $5.08 _{ - 0.28 } ^ { + 0.33 }$ & $0.34 _{ - 0.24 } ^ { + 0.29 }$ & $24.45 _{ - 0.47 } ^ { + 0.5 }$ & 9.178 &  & NRES (2) & Gemini &  \\
235943205 & TYC 4588-00127-1 & $1827.0267 _{ - 0.004 } ^ { + 0.0034 }$ & $121.3394 _{ - 0.0063 } ^ { + 0.0065 }$ & $0.0402 _{ - 0.0016 } ^ { + 0.0019 }$ & $4.2 _{ - 0.25 } ^ { + 0.29 }$ & $0.4 _{ - 0.27 } ^ { + 0.28 }$ & $6.37 _{ - 0.2 } ^ { + 0.3 }$ & 11.076 &  & NRES (1);SOPHIE (2) &  &  \\
237201858 & TYC 4452-00759-1 & $1811.5032 _{ - 0.0069 } ^ { + 0.0067 }$ & $129.7 _{ - 41.5 } ^ { + 146.8 }(s)$ & $0.0258 _{ - 0.0013 } ^ { + 0.0015 }$ & $4.12 _{ - 0.27 } ^ { + 0.3 }$ & $0.4 _{ - 0.28 } ^ { + 0.31 }$ & $10.94 _{ - 0.37 } ^ { + 0.53 }$ & 10.344 &  & NRES (1) &  &  \\
243187830* & HIP 5286 & $1783.7671 _{ - 0.0017 } ^ { + 0.0019 }$ & $4.05 _{ - 1.53 } ^ { + 9.21 }(s)$ & $0.0268 _{ - 0.0015 } ^ { + 0.0027 }$ & $2.06 _{ - 0.17 } ^ { + 0.23 }$ & $0.47 _{ - 0.32 } ^ { + 0.34 }$ & $2.02 _{ - 0.12 } ^ { + 0.15 }$ & 8.407 & SBIG (1) &  &  &  \\
243417115 & TYC 8262-02120-1 & $1614.4796 _{ - 0.0028 } ^ { + 0.0022 }$ & $1.81 _{ - 0.73 } ^ { + 3.45 }(s)$ & $0.0523 _{ - 0.0035 } ^ { + 0.005 }$ & $5.39 _{ - 0.47 } ^ { + 0.64 }$ & $0.47 _{ - 0.33 } ^ { + 0.34 }$ & $2.03 _{ - 0.16 } ^ { + 0.23 }$ & 11.553 &  &  &  &  \\
256429408 & TYC 4462-01942-1 & $1962.16 _{ - 0.0022 } ^ { + 0.0023 }$ & $382.0 _{ - 132.0 } ^ { + 265.0 }(s)$ & $0.03582 _{ - 0.00086 } ^ { + 0.00094 }$ & $6.12 _{ - 0.29 } ^ { + 0.3 }$ & $0.51 _{ - 0.36 } ^ { + 0.18 }$ & $16.96 _{ - 0.2 } ^ { + 0.24 }$ & 8.898 &  &  &  &  \\
264544388* & TYC 4607-01275-1 & $1824.8438 _{ - 0.0076 } ^ { + 0.0078 }$ & $7030.0 _{ - 6260.0 } ^ { + 3330.0 }(s)$ & $0.0288 _{ - 0.0029 } ^ { + 0.0018 }$ & $4.58 _{ - 0.43 } ^ { + 0.35 }$ & $0.936 _{ - 0.363 } ^ { + 0.011 }$ & $19.13 _{ - 1.35 } ^ { + 0.84 }$ & 8.758 &  & NRES (1) &  &  \\
264766922 & TYC 8565-01780-1 & $1538.69518 _{ - 0.00091 } ^ { + 0.00091 }$ & $3.28 _{ - 0.94 } ^ { + 1.25 }(s)$ & $0.0933 _{ - 0.0063 } ^ { + 0.0176 }$ & $16.95 _{ - 1.33 } ^ { + 3.19 }$ & $0.908 _{ - 0.039 } ^ { + 0.048 }$ & $2.73 _{ - 0.11 } ^ { + 0.11 }$ & 10.747 &  &  &  &  \\
26547036* & TYC 3921-01563-1 & $1712.30464 _{ - 0.00041 } ^ { + 0.0004 }$ & $73.0 _{ - 13.6 } ^ { + 16.5 }(s)$ & $0.10034 _{ - 0.0007 } ^ { + 0.00078 }$ & $11.75 _{ - 0.59 } ^ { + 0.58 }$ & $0.17 _{ - 0.12 } ^ { + 0.11 }$ & $8.681 _{ - 0.049 } ^ { + 0.052 }$ & 9.849 &  & NRES (4) & Gemini &  \\
267542728$\dagger$ & TYC 4583-01499-1 & $1708.4956 _{ - 0.0073 } ^ { + 0.0085 }$ & $39.7382 _{ - 0.0023 } ^ { + 0.0023 }$ & $0.03267 _{ - 0.00089 } ^ { + 0.00175 }$ & $18.46 _{ - 0.94 } ^ { + 1.14 }$ & $0.38 _{ - 0.26 } ^ { + 0.27 }$ & $24.16 _{ - 0.39 } ^ { + 0.45 }$ & 11.474 &  &  &  & EB from HIRES RVs. \\
270371513$\dagger$ & HIP 10047 & $1426.2967 _{ - 0.0023 } ^ { + 0.002 }$ & $0.39 _{ - 0.17 } ^ { + 1.79 }(s)$ & $0.024 _{ - 0.0015 } ^ { + 0.0032 }$ & $4.8 _{ - 0.38 } ^ { + 0.64 }$ & $0.5 _{ - 0.34 } ^ { + 0.39 }$ & $1.93 _{ - 0.16 } ^ { + 0.19 }$ & 6.98515 &  & MINERVA (20) &  & SB 2 from MINERVA observations.  \\
274599700 & TWOMASS 17011885+5131455 & $2002.1202 _{ - 0.0024 } ^ { + 0.0024 }$ & $32.9754 _{ - 0.005 } ^ { + 0.005 }$ & $0.0847 _{ - 0.0021 } ^ { + 0.0018 }$ & $13.25 _{ - 0.83 } ^ { + 0.83 }$ & $0.37 _{ - 0.24 } ^ { + 0.19 }$ & $8.2 _{ - 0.18 } ^ { + 0.21 }$ & 12.411 &  &  &  &  \\
278990954 & TYC 8548-00717-1 & $1650.0191 _{ - 0.0086 } ^ { + 0.0105 }$ & $18.45 _{ - 8.66 } ^ { + 230.7 }(s)$ & $0.034 _{ - 0.0024 } ^ { + 0.0115 }$ & $9.65 _{ - 0.92 } ^ { + 3.13 }$ & $0.58 _{ - 0.4 } ^ { + 0.36 }$ & $10.62 _{ - 0.66 } ^ { + 2.46 }$ & 10.749 &  &  &  &  \\
280865159* & TYC 9384-01533-1 & $1387.0749 _{ - 0.0045 } ^ { + 0.0044 }$ & $1045.0 _{ - 249.0 } ^ { + 536.0 }(s)$ & $0.0406 _{ - 0.0011 } ^ { + 0.0014 }$ & $4.75 _{ - 0.26 } ^ { + 0.28 }$ & $0.35 _{ - 0.24 } ^ { + 0.23 }$ & $19.08 _{ - 0.32 } ^ { + 0.36 }$ & 11.517 &  &  & Gemini &  \\
284361752 & TYC 3924-01678-1 & $2032.093 _{ - 0.0078 } ^ { + 0.008 }$ & $140.6 _{ - 46.6 } ^ { + 159.1 }(s)$ & $0.0259 _{ - 0.0014 } ^ { + 0.0017 }$ & $3.62 _{ - 0.26 } ^ { + 0.31 }$ & $0.4 _{ - 0.27 } ^ { + 0.34 }$ & $8.98 _{ - 0.66 } ^ { + 0.86 }$ & 10.221 &  &  &  &  \\
288240183 & TYC 4634-01225-1 & $1896.941 _{ - 0.0051 } ^ { + 0.0047 }$ & $119.0502 _{ - 0.0091 } ^ { + 0.0089 }$ & $0.02826 _{ - 0.00089 } ^ { + 0.00119 }$ & $4.28 _{ - 0.35 } ^ { + 0.36 }$ & $0.55 _{ - 0.37 } ^ { + 0.25 }$ & $17.49 _{ - 0.36 } ^ { + 0.6 }$ & 9.546 &  &  &  &  \\
29169215 & TWOMASS 09011787+4727085 & $1872.5047 _{ - 0.0032 } ^ { + 0.0036 }$ & $14.89 _{ - 6.12 } ^ { + 24.84 }(s)$ & $0.0403 _{ - 0.0025 } ^ { + 0.0033 }$ & $3.28 _{ - 0.37 } ^ { + 0.45 }$ & $0.44 _{ - 0.3 } ^ { + 0.33 }$ & $3.56 _{ - 0.21 } ^ { + 0.32 }$ & 11.828 &  &  &  &  \\
293649602 & TYC 8103-00266-1 & $1511.2109 _{ - 0.004 } ^ { + 0.0037 }$ & $12.85 _{ - 5.34 } ^ { + 42.21 }(s)$ & $0.04 _{ - 0.0024 } ^ { + 0.0039 }$ & $4.66 _{ - 0.36 } ^ { + 0.5 }$ & $0.5 _{ - 0.35 } ^ { + 0.34 }$ & $4.1 _{ - 0.31 } ^ { + 0.56 }$ & 10.925 &  &  &  &  \\
296737508 & TYC 5472-01060-1 & $1538.0036 _{ - 0.0015 } ^ { + 0.0016 }$ & $18.27 _{ - 5.06 } ^ { + 17.45 }(s)$ & $0.0425 _{ - 0.0014 } ^ { + 0.0019 }$ & $5.33 _{ - 0.22 } ^ { + 0.27 }$ & $0.44 _{ - 0.3 } ^ { + 0.26 }$ & $5.13 _{ - 0.13 } ^ { + 0.15 }$ & 9.772 & Sinistro (1) & NRES (1);MINERVA (1) & Gemini &  \\
298663873 & TYC 3913-01781-1 & $1830.76819 _{ - 0.00099 } ^ { + 0.00099 }$ & $479.9 _{ - 89.4 } ^ { + 109.4 }(s)$ & $0.06231 _{ - 0.00034 } ^ { + 0.00045 }$ & $11.07 _{ - 0.57 } ^ { + 0.57 }$ & $0.16 _{ - 0.11 } ^ { + 0.13 }$ & $23.99 _{ - 0.093 } ^ { + 0.1 }$ & 9.162 &  & NRES (2) & Gemini & Dalba et al. (in prep) \\
303050301 & TYC 6979-01108-1 & $1366.1301 _{ - 0.0022 } ^ { + 0.0023 }$ & $281.0 _{ - 170.0 } ^ { + 264.0 }(s)$ & $0.0514 _{ - 0.0027 } ^ { + 0.0018 }$ & $4.85 _{ - 0.32 } ^ { + 0.32 }$ & $0.73 _{ - 0.48 } ^ { + 0.1 }$ & $7.91 _{ - 0.31 } ^ { + 0.36 }$ & 10.048 &  & NRES (1) & Gemini &  \\
303317324 & TYC 6983-00438-1 & $1365.1845 _{ - 0.0023 } ^ { + 0.0028 }$ & $69.0 _{ - 25.5 } ^ { + 78.1 }(s)$ & $0.0365 _{ - 0.0013 } ^ { + 0.0016 }$ & $2.88 _{ - 0.3 } ^ { + 0.31 }$ & $0.39 _{ - 0.26 } ^ { + 0.32 }$ & $5.78 _{ - 0.18 } ^ { + 0.24 }$ & 10.799 &  &  &  &  \\
\hline
\end{tabular}}
\caption{\emph{Note} -- Candidates that have become TOIs following the PHT discovery are marked with an asterisk (*). The `s' following the orbital period indicates that the candidates is a single transit event. The ground-based follow-up observations are summarized in columns 10-12, where the bracketed numbers correspond the number of epochs obtained with each instrument.  See Section~\ref{sec:follow_up} for description of each instrument. The $\dagger$ symbol indicates candidates that have been shown to be astrophysical false  positives based on the ground based follow-up observations.}
\label{tab:PHT-caniddates}
\end{table}
\end{landscape}

\begin{landscape}
\begin{table}
\addtocounter{table}{-1}
\resizebox{1.31\textwidth}{!}{
\begin{tabular}{ccccccccccccc}
\textbf{TIC} & \textbf{Other} & \textbf{Epoch} & \textbf{Period}  & \textbf{$R_{pl}$/$R_{\odot}$} & \textbf{$R_{pl}$} & \textbf{Impact} & \textbf{Duration} & \textbf{$V_{mag}$} & \textbf{Photometry} & \textbf{Spectroscopy} & \textbf{Speckle} & \textbf{Comment} \\
  & \textbf{Name} & \textbf{(\textcolor{black}{BJD - 2457000})} & \textbf{(days)}  & & $(R_{\oplus})$  & \textbf{Parameter} & \textbf{(hours)} &   &   &   &   &   \\
\hline
303586471$\dagger$ & HIP 115828 & $1363.7692 _{ - 0.0033 } ^ { + 0.0027 }$ & $13.85 _{ - 4.19 } ^ { + 18.2 }(s)$ & $0.0214 _{ - 0.001 } ^ { + 0.0014 }$ & $2.52 _{ - 0.16 } ^ { + 0.2 }$ & $0.4 _{ - 0.27 } ^ { + 0.33 }$ & $4.23 _{ - 0.19 } ^ { + 0.16 }$ & 8.27 &  & MINERVA (11) &  & SB 2 from MINERVA observations.  \\
304142124* & HIP 53719 & $1585.28023 _{ - 0.0008 } ^ { + 0.0008 }$ & $42.8 _{ - 10.0 } ^ { + 18.2 }(s)$ & $0.04311 _{ - 0.00093 } ^ { + 0.00153 }$ & $4.1 _{ - 0.23 } ^ { + 0.24 }$ & $0.33 _{ - 0.21 } ^ { + 0.21 }$ & $5.66 _{ - 0.067 } ^ { + 0.09 }$ & 8.62 &  & NRES (1);MINERVA (4) &  & Confirmed planet \citep{diaz2020} \\
304339227 & TYC 9290-01087-1 & $1673.3242 _{ - 0.009 } ^ { + 0.0128 }$ & $111.9 _{ - 72.2 } ^ { + 4844.1 }(s)$ & $0.0253 _{ - 0.0024 } ^ { + 0.0481 }$ & $3.27 _{ - 0.61 } ^ { + 5.72 }$ & $0.67 _{ - 0.47 } ^ { + 0.36 }$ & $7.44 _{ - 0.86 } ^ { + 2.84 }$ & 9.169 &  &  &  &  \\
307958020 & TYC 4191-00309-1 & $1864.82 _{ - 0.014 } ^ { + 0.013 }$ & $169.0 _{ - 107.0 } ^ { + 10194.0 }(s)$ & $0.0223 _{ - 0.0022 } ^ { + 0.0543 }$ & $3.92 _{ - 0.52 } ^ { + 9.27 }$ & $0.71 _{ - 0.53 } ^ { + 0.33 }$ & $12.48 _{ - 1.1 } ^ { + 5.41 }$ & 9.017 &  &  &  &  \\
308301091 & TYC 2081-01273-1 & $2030.3691 _{ - 0.0024 } ^ { + 0.0026 }$ & $29.24 _{ - 8.49 } ^ { + 22.46 }(s)$ & $0.0362 _{ - 0.0013 } ^ { + 0.0014 }$ & $5.41 _{ - 0.34 } ^ { + 0.35 }$ & $0.35 _{ - 0.25 } ^ { + 0.29 }$ & $6.57 _{ - 0.14 } ^ { + 0.19 }$ & 10.273 &  &  &  &  \\
313006381 & HIP 45012 & $1705.687 _{ - 0.0081 } ^ { + 0.0045 }$ & $21.56 _{ - 8.9 } ^ { + 54.15 }(s)$ & $0.0261 _{ - 0.0017 } ^ { + 0.0027 }$ & $2.34 _{ - 0.2 } ^ { + 0.27 }$ & $0.45 _{ - 0.3 } ^ { + 0.38 }$ & $3.85 _{ - 0.51 } ^ { + 0.31 }$ & 9.39 &  &  &  &  \\
323295479* & TYC 9506-01881-1 & $1622.9258 _{ - 0.00083 } ^ { + 0.00087 }$ & $117.8 _{ - 25.8 } ^ { + 30.9 }(s)$ & $0.0981 _{ - 0.0021 } ^ { + 0.0023 }$ & $11.35 _{ - 0.67 } ^ { + 0.66 }$ & $0.839 _{ - 0.024 } ^ { + 0.019 }$ & $6.7 _{ - 0.14 } ^ { + 0.15 }$ & 10.595 &  &  &  &  \\
328933398.01* & TYC 4634-01435-1 & $1880.9878 _{ - 0.0039 } ^ { + 0.0042 }$ & $24.9335 _{ - 0.0046 } ^ { + 0.005 }$ & $0.0437 _{ - 0.0022 } ^ { + 0.0023 }$ & $4.62 _{ - 0.32 } ^ { + 0.33 }$ & $0.38 _{ - 0.25 } ^ { + 0.27 }$ & $5.02 _{ - 0.22 } ^ { + 0.27 }$ & 11.215 &  &  &  & Potential multi-planet system. \\
328933398.02* & TYC 4634-01435-1 & $1848.6557 _{ - 0.0053 } ^ { + 0.0072 }$ & $50.5 _{ - 22.4 } ^ { + 77.1 }(s)$ & $0.0296 _{ - 0.0028 } ^ { + 0.0033 }$ & $3.14 _{ - 0.33 } ^ { + 0.39 }$ & $0.41 _{ - 0.28 } ^ { + 0.35 }$ & $5.99 _{ - 0.8 } ^ { + 0.77 }$ & 11.215 &  &  &  &  \\
331644554 & TYC 3609-00469-1 & $1757.0354 _{ - 0.0031 } ^ { + 0.0033 }$ & $947.0 _{ - 215.0 } ^ { + 274.0 }(s)$ & $0.12 _{ - 0.025 } ^ { + 0.021 }$ & $21.84 _{ - 4.57 } ^ { + 3.86 }$ & $1.018 _{ - 0.036 } ^ { + 0.028 }$ & $10.93 _{ - 0.34 } ^ { + 0.35 }$ & 9.752 &  &  &  &  \\
332657786 & TWOMASS 09595797-1609323 & $1536.7659 _{ - 0.0015 } ^ { + 0.0015 }$ & $63.76 _{ - 9.52 } ^ { + 11.13 }(s)$ & $0.14961 _{ - 0.00064 } ^ { + 0.00029 }$ & $3.83 _{ - 0.12 } ^ { + 0.12 }$ & $0.059 _{ - 0.041 } ^ { + 0.064 }$ & $3.333 _{ - 0.095 } ^ { + 0.096 }$ & 15.99 &  &  &  &  \\
336075472 & TYC 3526-00332-1 & $2028.1762 _{ - 0.0043 } ^ { + 0.0037 }$ & $61.9 _{ - 24.0 } ^ { + 95.6 }(s)$ & $0.0402 _{ - 0.0022 } ^ { + 0.0033 }$ & $3.09 _{ - 0.34 } ^ { + 0.4 }$ & $0.43 _{ - 0.29 } ^ { + 0.32 }$ & $5.39 _{ - 0.23 } ^ { + 0.37 }$ & 11.842 &  &  &  &  \\
349488688.01 & TYC 1529-00224-1 & $1994.283 _{ - 0.0038 } ^ { + 0.0033 }$ & $11.6254 _{ - 0.005 } ^ { + 0.0052 }$ & $0.02195 _{ - 0.00096 } ^ { + 0.00122 }$ & $3.44 _{ - 0.18 } ^ { + 0.21 }$ & $0.39 _{ - 0.27 } ^ { + 0.3 }$ & $5.58 _{ - 0.15 } ^ { + 0.18 }$ & 8.855 &  & NRES (2);SOPHIE (2) &  & Potential multi-planet system. \\
349488688.02 & TYC 1529-00224-1 & $2002.77063 _{ - 0.00097 } ^ { + 0.00103 }$ & $15.35 _{ - 1.94 } ^ { + 4.15 }(s)$ & $0.03688 _{ - 0.00067 } ^ { + 0.00069 }$ & $5.78 _{ - 0.18 } ^ { + 0.18 }$ & $0.24 _{ - 0.16 } ^ { + 0.21 }$ & $6.291 _{ - 0.058 } ^ { + 0.074 }$ & 8.855 &  & NRES (2);SOPHIE (2) &  &  \\
356700488* & TYC 4420-01295-1 & $1756.638 _{ - 0.013 } ^ { + 0.011 }$ & $184.5 _{ - 64.7 } ^ { + 333.1 }(s)$ & $0.0173 _{ - 0.0011 } ^ { + 0.0015 }$ & $2.92 _{ - 0.2 } ^ { + 0.28 }$ & $0.44 _{ - 0.3 } ^ { + 0.34 }$ & $11.76 _{ - 0.65 } ^ { + 1.03 }$ & 8.413 &  &  &  &  \\
356710041* & TYC 1993-00419-1 & $1932.2939 _{ - 0.0019 } ^ { + 0.0019 }$ & $29.6 _{ - 14.0 } ^ { + 19.0 }(s)$ & $0.0496 _{ - 0.0021 } ^ { + 0.0011 }$ & $14.82 _{ - 0.85 } ^ { + 0.84 }$ & $0.66 _{ - 0.42 } ^ { + 0.11 }$ & $12.76 _{ - 0.24 } ^ { + 0.24 }$ & 9.646 &  &  &  &  \\
369532319 & TYC 2743-01716-1 & $1755.8158 _{ - 0.006 } ^ { + 0.0051 }$ & $35.4 _{ - 12.0 } ^ { + 51.6 }(s)$ & $0.0316 _{ - 0.0023 } ^ { + 0.0028 }$ & $3.43 _{ - 0.3 } ^ { + 0.37 }$ & $0.41 _{ - 0.29 } ^ { + 0.34 }$ & $5.5 _{ - 0.32 } ^ { + 0.32 }$ & 10.594 &  &  & Gemini &  \\
369779127 & TYC 9510-00090-1 & $1643.9403 _{ - 0.0046 } ^ { + 0.0058 }$ & $9.93 _{ - 3.38 } ^ { + 19.74 }(s)$ & $0.0288 _{ - 0.0015 } ^ { + 0.0033 }$ & $4.89 _{ - 0.31 } ^ { + 0.56 }$ & $0.46 _{ - 0.31 } ^ { + 0.33 }$ & $5.64 _{ - 0.38 } ^ { + 0.33 }$ & 9.279 &  &  &  &  \\
384159646* & TYC 9454-00957-1 & $1630.39405 _{ - 0.00079 } ^ { + 0.00079 }$ & $11.68 _{ - 2.75 } ^ { + 4.21 }(s)$ & $0.0658 _{ - 0.0012 } ^ { + 0.0011 }$ & $9.87 _{ - 0.45 } ^ { + 0.44 }$ & $0.27 _{ - 0.18 } ^ { + 0.21 }$ & $5.152 _{ - 0.069 } ^ { + 0.087 }$ & 10.158 & SBIG (1) & NRES (8);MINERVA (6) & Gemini &  \\
385557214 & TYC 1807-00046-1 & $1791.58399 _{ - 0.00068 } ^ { + 0.0007 }$ & $5.62451 _{ - 0.0004 } ^ { + 0.00043 }$ & $0.096 _{ - 0.019 } ^ { + 0.032 }$ & $8.32 _{ - 2.06 } ^ { + 2.77 }$ & $0.95 _{ - 0.075 } ^ { + 0.053 }$ & $1.221 _{ - 0.094 } ^ { + 0.058 }$ & 10.856 &  &  &  &  \\
388134787 & TYC 4260-00427-1 & $1811.034 _{ - 0.015 } ^ { + 0.017 }$ & $246.0 _{ - 127.0 } ^ { + 6209.0 }(s)$ & $0.0265 _{ - 0.0024 } ^ { + 0.023 }$ & $2.57 _{ - 0.28 } ^ { + 2.19 }$ & $0.55 _{ - 0.39 } ^ { + 0.44 }$ & $8.85 _{ - 1.13 } ^ { + 1.84 }$ & 10.95 &  & NRES (1) & Gemini &  \\
404518509 & HIP 16038 & $1431.2696 _{ - 0.0037 } ^ { + 0.0035 }$ & $26.83 _{ - 9.46 } ^ { + 56.14 }(s)$ & $0.0259 _{ - 0.0013 } ^ { + 0.0022 }$ & $2.94 _{ - 0.21 } ^ { + 0.29 }$ & $0.47 _{ - 0.31 } ^ { + 0.34 }$ & $5.02 _{ - 0.23 } ^ { + 0.28 }$ & 9.17 &  &  &  &  \\
408636441* & TYC 4266-00736-1 & $1745.4668 _{ - 0.0016 } ^ { + 0.0015 }$ & $37.695 _{ - 0.0034 } ^ { + 0.0033 }$ & $0.0485 _{ - 0.0019 } ^ { + 0.0023 }$ & $3.32 _{ - 0.16 } ^ { + 0.19 }$ & $0.39 _{ - 0.27 } ^ { + 0.29 }$ & $3.63 _{ - 0.1 } ^ { + 0.14 }$ & 11.93 & SBIG (1) &  & Gemini & Half of the period likely. \\
418255064 & TWOMASS 13063680-8037015 & $1629.3304 _{ - 0.0018 } ^ { + 0.0018 }$ & $25.37 _{ - 7.06 } ^ { + 15.41 }(s)$ & $0.0732 _{ - 0.0029 } ^ { + 0.0031 }$ & $5.57 _{ - 0.36 } ^ { + 0.38 }$ & $0.37 _{ - 0.25 } ^ { + 0.25 }$ & $3.83 _{ - 0.13 } ^ { + 0.14 }$ & 12.478 & SBIG (1) &  & Gemini &  \\
420645189$\dagger$ & TYC 4508-00478-1 & $1837.4767 _{ - 0.0018 } ^ { + 0.0017 }$ & $250.2 _{ - 66.6 } ^ { + 99.4 }(s)$ & $0.0784 _{ - 0.0033 } ^ { + 0.0046 }$ & $8.82 _{ - 0.55 } ^ { + 0.7 }$ & $0.892 _{ - 0.026 } ^ { + 0.028 }$ & $6.95 _{ - 0.27 } ^ { + 0.3 }$ & 10.595 &  & MINERVA (1) &  & SB 2 from MINERVA observations.  \\
422914082 & TYC 0046-00133-1 & $1431.5538 _{ - 0.0014 } ^ { + 0.0017 }$ & $12.91 _{ - 3.91 } ^ { + 8.97 }(s)$ & $0.0418 _{ - 0.0015 } ^ { + 0.0016 }$ & $3.96 _{ - 0.32 } ^ { + 0.35 }$ & $0.36 _{ - 0.25 } ^ { + 0.28 }$ & $4.07 _{ - 0.09 } ^ { + 0.126 }$ & 11.026 & Sinistro (1) & NRES (1) &  &  \\
427344083 & TWOMASS 22563609+7040518 & $1961.8967 _{ - 0.0031 } ^ { + 0.0036 }$ & $7.77 _{ - 5.6 } ^ { + 9.65 }(s)$ & $0.107 _{ - 0.016 } ^ { + 0.025 }$ & $12.27 _{ - 1.87 } ^ { + 2.9 }$ & $0.834 _{ - 0.484 } ^ { + 0.094 }$ & $2.88 _{ - 0.3 } ^ { + 0.42 }$ & 13.404 &  &  &  &  \\
436873727 & HIP 13224 & $1803.83679 _{ - 0.00058 } ^ { + 0.00056 }$ & $19.26 _{ - 5.95 } ^ { + 6.73 }(s)$ & $0.05246 _{ - 0.00061 } ^ { + 0.00059 }$ & $10.02 _{ - 0.43 } ^ { + 0.41 }$ & $0.767 _{ - 0.057 } ^ { + 0.038 }$ & $5.462 _{ - 0.081 } ^ { + 0.074 }$ & 7.51 &  &  &  &  \\ 
441642457* & TYC 3858-00452-1 & $1745.5102 _{ - 0.0108 } ^ { + 0.0097 }$ & $79.8072 _{ - 0.0071 } ^ { + 0.0076 }$ & $0.0281 _{ - 0.0024 } ^ { + 0.0033 }$ & $3.55 _{ - 0.34 } ^ { + 0.46 }$ & $0.934 _{ - 0.023 } ^ { + 0.026 }$ & $6.9 _{ - 0.39 } ^ { + 0.6 }$ & 9.996 &  &  &  &  \\
441765914* & TWOMASS 17253007+7552562 & $1769.6154 _{ - 0.0058 } ^ { + 0.0093 }$ & $161.6 _{ - 58.2 } ^ { + 1460.1 }(s)$ & $0.0411 _{ - 0.0024 } ^ { + 0.0119 }$ & $3.6 _{ - 0.3 } ^ { + 1.01 }$ & $0.45 _{ - 0.32 } ^ { + 0.48 }$ & $7.44 _{ - 0.36 } ^ { + 1.08 }$ & 11.638 &  &  &  &  \\
452920657 & TWOMASS 00332018+5906355 & $1810.5765 _{ - 0.0031 } ^ { + 0.003 }$ & $53.2 _{ - 29.0 } ^ { + 34.3 }(s)$ & $0.135 _{ - 0.016 } ^ { + 0.012 }$ & $9.71 _{ - 1.16 } ^ { + 0.9 }$ & $0.73 _{ - 0.48 } ^ { + 0.11 }$ & $4.6 _{ - 0.26 } ^ { + 0.29 }$ & 14.167 & SBIG (1) &  &  &  \\
455737331 & TYC 2779-00785-1 & $1780.7084 _{ - 0.008 } ^ { + 0.0073 }$ & $50.4 _{ - 17.6 } ^ { + 75.0 }(s)$ & $0.0257 _{ - 0.0016 } ^ { + 0.002 }$ & $3.05 _{ - 0.24 } ^ { + 0.29 }$ & $0.43 _{ - 0.29 } ^ { + 0.33 }$ & $6.6 _{ - 0.43 } ^ { + 0.5 }$ & 10.189 & SBIG (1) &  & Gemini &  \\
456909420 & TYC 1208-01094-1 & $1779.4109 _{ - 0.0026 } ^ { + 0.0022 }$ & $5.78 _{ - 5.29 } ^ { + 5.95 }(s)$ & $0.078 _{ - 0.031 } ^ { + 0.045 }$ & $9.15 _{ - 3.61 } ^ { + 5.27 }$ & $0.973 _{ - 0.495 } ^ { + 0.063 }$ & $1.73 _{ - 0.27 } ^ { + 0.28 }$ & 10.941 &  &  &  &  \\
458451774 & TWOMASS 12551793+4431260 & $1917.1875 _{ - 0.0019 } ^ { + 0.0019 }$ & $12.39 _{ - 6.34 } ^ { + 83.97 }(s)$ & $0.0752 _{ - 0.0054 } ^ { + 0.0211 }$ & $3.33 _{ - 0.26 } ^ { + 0.92 }$ & $0.61 _{ - 0.43 } ^ { + 0.32 }$ & $2.08 _{ - 0.19 } ^ { + 0.59 }$ & 13.713 &  &  &  &  \\
48018596 & TYC 3548-00800-1 & $1713.4514 _{ - 0.0063 } ^ { + 0.0046 }$ & $100.1145 _{ - 0.0018 } ^ { + 0.0021 }$ & $0.049 _{ - 0.0081 } ^ { + 0.018 }$ & $7.88 _{ - 1.33 } ^ { + 2.9 }$ & $0.984 _{ - 0.028 } ^ { + 0.027 }$ & $2.83 _{ - 0.26 } ^ { + 0.29 }$ & 9.595 &  & NRES (1) & Gemini &  \\
53309262 & TWOMASS 07475406+5741549 & $1863.1133 _{ - 0.0064 } ^ { + 0.0061 }$ & $294.8 _{ - 96.0 } ^ { + 327.0 }(s)$ & $0.1239 _{ - 0.0075 } ^ { + 0.0098 }$ & $5.38 _{ - 0.36 } ^ { + 0.46 }$ & $0.46 _{ - 0.31 } ^ { + 0.28 }$ & $6.74 _{ - 0.45 } ^ { + 0.62 }$ & 15.51 &  &  &  &  \\
53843023 & TYC 6956-00758-1 & $1328.0335 _{ - 0.0054 } ^ { + 0.0057 }$ & $202.0 _{ - 189.0 } ^ { + 272.0 }(s)$ & $0.058 _{ - 0.02 } ^ { + 0.056 }$ & $5.14 _{ - 1.77 } ^ { + 4.99 }$ & $0.962 _{ - 0.597 } ^ { + 0.083 }$ & $4.25 _{ - 0.72 } ^ { + 0.66 }$ & 11.571 &  &  &  &  \\
55525572* & TYC 8876-01059-1 & $1454.6713 _{ - 0.0066 } ^ { + 0.0065 }$ & $83.8951 _{ - 0.004 } ^ { + 0.004 }$ & $0.0343 _{ - 0.001 } ^ { + 0.0021 }$ & $7.31 _{ - 0.46 } ^ { + 0.56 }$ & $0.43 _{ - 0.29 } ^ { + 0.31 }$ & $13.54 _{ - 0.3 } ^ { + 0.51 }$ & 10.358 &  & CHIRON (5) & Gemini & Confirmed planet \citep{2020eisner} \\
63698669* & TYC 6993-00729-1 & $1364.6226 _{ - 0.0074 } ^ { + 0.0067 }$ & $73.6 _{ - 26.8 } ^ { + 133.6 }(s)$ & $0.0248 _{ - 0.0019 } ^ { + 0.0023 }$ & $2.15 _{ - 0.2 } ^ { + 0.25 }$ & $0.42 _{ - 0.29 } ^ { + 0.35 }$ & $5.63 _{ - 0.32 } ^ { + 0.57 }$ & 10.701 & SBIG (1) &  &  &  \\
70887357* & TYC 5883-01412-1 & $1454.3341 _{ - 0.0016 } ^ { + 0.0015 }$ & $56.1 _{ - 15.3 } ^ { + 18.8 }(s)$ & $0.0605 _{ - 0.0027 } ^ { + 0.0027 }$ & $12.84 _{ - 0.86 } ^ { + 0.9 }$ & $0.917 _{ - 0.028 } ^ { + 0.016 }$ & $7.29 _{ - 0.18 } ^ { + 0.19 }$ & 9.293 &  &  &  &  \\
7422496$\dagger$ & HIP 25359 & $1470.3625 _{ - 0.0031 } ^ { + 0.0023 }$ & $61.4 _{ - 16.7 } ^ { + 49.0 }(s)$ & $0.0255 _{ - 0.001 } ^ { + 0.0011 }$ & $2.44 _{ - 0.15 } ^ { + 0.16 }$ & $0.37 _{ - 0.25 } ^ { + 0.29 }$ & $5.89 _{ - 0.15 } ^ { + 0.15 }$ & 9.36 &  & MINERVA (4) &  & SB 2 from MINERVA observations.  \\
82452140 & TYC 3076-00921-1 & $1964.292 _{ - 0.011 } ^ { + 0.011 }$ & $21.1338 _{ - 0.0052 } ^ { + 0.0066 }$ & $0.0266 _{ - 0.0019 } ^ { + 0.0027 }$ & $2.95 _{ - 0.25 } ^ { + 0.34 }$ & $0.42 _{ - 0.29 } ^ { + 0.36 }$ & $5.87 _{ - 0.62 } ^ { + 0.94 }$ & 10.616 &  &  &  &  \\
88840705 & TYC 3091-00808-1 & $2026.6489 _{ - 0.001 } ^ { + 0.001 }$ & $260.6 _{ - 87.6 } ^ { + 142.2 }(s)$ & $0.109 _{ - 0.023 } ^ { + 0.027 }$ & $9.98 _{ - 2.28 } ^ { + 2.75 }$ & $1.001 _{ - 0.042 } ^ { + 0.037 }$ & $4.72 _{ - 0.13 } ^ { + 0.15 }$ & 9.443 &  &  &  &  \\
91987762* & HIP 47288 & $1894.25381 _{ - 0.00051 } ^ { + 0.00047 }$ & $10.51 _{ - 3.48 } ^ { + 3.67 }(s)$ & $0.05459 _{ - 0.00106 } ^ { + 0.00097 }$ & $9.56 _{ - 0.56 } ^ { + 0.52 }$ & $0.771 _{ - 0.062 } ^ { + 0.033 }$ & $4.342 _{ - 0.073 } ^ { + 0.063 }$ & 7.87 &  & NRES (4) & Gemini &  \\
95768667 & TYC 1434-00331-1 & $1918.3318 _{ - 0.0093 } ^ { + 0.0079 }$ & $26.9 _{ - 12.4 } ^ { + 72.3 }(s)$ & $0.0282 _{ - 0.0022 } ^ { + 0.0031 }$ & $3.54 _{ - 0.32 } ^ { + 0.43 }$ & $0.48 _{ - 0.33 } ^ { + 0.35 }$ & $5.4 _{ - 0.64 } ^ { + 0.76 }$ & 10.318 &  &  &  &  \\
\hline
\end{tabular}}
\caption{\textbf{Properties of PHT candidates (continued)}}
\label{tab:PHT-caniddates2}
\end{table}
\end{landscape}

\section{Conclusion}
\label{sec:condlusion}

We present the results from the analysis of the first 26  \tess\ sectors. The outlined citizen science approach engages over 22 thousand registered citizen scientists who completed 12,617,038 classifications from December 2018 through August 2020 for the sectors observed during the first two years of the \tess\ mission. We applied a systematic search for planetary candidates using visual vetting by multiple volunteers to identify \tess\ targets that are most likely to host a planet. Between 8 and 15 volunteers have inspected each \tess\ light curve and marked times of transit-like events using the PHT online interface. For each light curve, the markings from all the volunteers who saw that target were combined using an unsupervised machine learning method, known as DBSCAN, in order to identify likely transit-like events. Each of these identified events was given a transit score based on the number of volunteers who identified a given event and on the user weighting of each of those volunteers. Individual user weights were calculated based on the user's ability to identify simulated transit events, injected into real \tess\ light curves, that are displayed on the PHT site alongside of the real data. The transit scores were then used to generate a ranked list of candidates that range from most likely to least likely to host a planet candidate. The top 500 highest ranked candidates were further vetted by the PHT science team. This stage of vetting primarily made use of the open source {\sc latte} \citep{LATTE2020} tool which generates a number of standard diagnostic plots that help identify promising candidates and weed out false positive signals. 

On average we found around three high priority candidates per sector which were followed up using ground based telescopes, where possible. To date, PHT has statistically confirmed one planet, TOI-813 \citep{2020eisner}: a Saturn-sized planet on an 84 day orbit around a subgiant host star. Other PHT identified planets listed in this paper are being followed up by other teams of astronomers, such as TOI-1899 (TIC 172370679) which was recently confirmed to be a warm Jupiter transiting an M-dwarf \citep{canas2020}. The remaining candidates outlined in this paper require further follow-up observations to confirm their planetary nature.

The sensitivity of our transit search effort was assessed using synthetic data, as well as the known TOI and TCE candidates flagged by the SPOC pipeline. For simulated planets (where simulated signals are injected into real \tess\ light curves) we have shown that the recovery efficiency of human vetting starts to decrease for transit-signals that have a SNR less than 7.5. The detection efficiency was further evaluated by the fractional recovery of the TOI and TCEs. We have shown that PHT is over 85 \% complete in the recovery of planets that have a radius greater than 4 $R_{\oplus}$, 51 \% complete for radii between 3 and 4 $R_{\oplus}$ and 49 \% complete for radii between 2 and 3 $R_{\oplus}$. Furthermore, we have shown that human vetting is not sensitive to the number of transits present in the light curve, meaning that they are equally likely to identify candidates on longer orbital periods as they are those with shorter orbital periods for periods greater than $\sim$ 1 day. Planets with periods shorter than around 1 day exhibit over 20 transits within one \tess\ sectors resulting in a decrease in identification by the volunteers. This is due to many volunteers only marking a random subset of these events, resulting in a lack of consensus on any given transit event and thus decreasing the overall transit score of these light curves. 

In addition to searching for signals due to transiting exoplanets, PHT provides a platform that can be used to identify other stellar phenomena that may otherwise be difficult to identify with automated pipelines. Such phenomena, including eclipsing binaries, multiple stellar systems, dwarf novae, and stellar flares are often mentioned on the PHT discussion forums where volunteers can use searchable hashtags and comments to bring these systems to the attention of other citizen scientists as well as the PHT science team. All of the eclipsing binaries identified on the site, for example, are being used and vetted by the \tess\ Eclipsing Binary Working Group (Prsa et al. in prep). Furthermore, we have investigated the nature of all of the targets that were identified as possible multiple stellar systems, as summarised in Table~\ref{tab:PHT-multis}.

Overall we have shown that large scale visual vetting can complement the findings \textcolor{black}{from the major \tess\ pipeline} by identifying longer period planets that may only exhibit a single transit event in their light curve, as well as in finding signals that are aperiodic or embedded in a strong varying stellar signal. The identification of planets around stars with variable signals allow us to potentially characterise the host-star (e.g., with asteroseismology or spot modulation). Additionally, the longer period planets are integral to our understanding of how planet systems form and evolve, as they allow us to investigate the evolution of planets that are farther away from their host star and therefore less dependent on stellar radiation. \textcolor{black}{While automated pipelines specifically designed to identify single transit events in the \tess\ data exist \citep[e.g., ][]{Gill2020}, neither their methodology nor the full list of their findings are yet publicly available and thus we are unable to compare results.} 

The planets that PHT finds have longer periods ($\gtrsim$ 27 d) than those found in \tess\ data using automated pipelines, and are more typical of the Kepler sample (25\% of Kepler confirmed planets have periods greater than 27 days\footnote{\url{https://exoplanetarchive.ipac.caltech.edu/}}). However, the Kepler planets are considerably fainter, and thus less amenable to ground-based follow-up or atmospheric characterisation from space (CHEOPS and JWST). Thus PHT helps to bridge the parameter spaces covered by these two missions, by identifying longer period planet candidates around bright, nearby stars, for which we can ultimately obtain precise planetary mass estimates. Although statistical characterisation of exo-planetary systems is no doubt important, precise mass measurements are key to developing our understanding of exoplanets and the physics which dictate their evolution. In particular, identification of this PHT sample provides follow-up targets to investigate the dependence of photo-evaporation on the mass of planets as well as on the planet radius, and will help our understanding of the photo-evaporation valley at longer orbital periods \citep{Owen2013}. 

PHT will continue to operate throughout the \tess\ extended mission, hopefully allowing us to identify even longer period planets as well as help verify some of the existing candidates with additional transits.  



\begin{table*}
\resizebox{0.95\textwidth}{!}{
\begin{tabular}{cccccccccc}
\textbf{TIC}  & \textbf{Period (days)}        & \textbf{Epoch (\textcolor{black}{BJD - 2457000})}      & \textbf{Depth (ppm)}  & \textbf{Comment}  \\
\hline
13968858      & $3.4850 \pm 0.001$            & $ 1684.780   \pm 0.005$            & 410000    & Candidate multiple system  \\
              & $1.4380 \pm 0.001$            & $ 1684.335   \pm 0.005$            & 50000     &   \\
35655828      & $ 8.073 \pm 0.01$             & $ 1550.94    \pm 0.01 $            & 23000     & Confirmed blend \\
              & $  1.220 \pm 0.001 $          & $ 1545.540   \pm 0.005 $           & 2800      &  \\
63291675      & $ 8.099  \pm 0.003 $          & $ 1685.1     \pm 0.01  $           & 60000     & Confirmed blend \\
              & $ 1.4635 \pm 0.0005 $         & $ 1683.8     \pm 0.1 $             & 7000      &  \\
63459761      & $4.3630 \pm 0.003 $           & $ 1714.350   \pm 0.005 $           & 160000    & Candidate multiple system   \\
              & $4.235 \pm  $ 0.005           & $ 1715.130   \pm 0.03$             & 35000     &   \\
104909909     & $1.3060 \pm 0.0001$           & $ 1684.470   \pm 0.005$            & 32000     & Candidate multiple system   \\
              & $2.5750 \pm 0.003$            & $ 1684.400   \pm 0.005$            & 65000     &   \\
115980439     & $ 4.615 \pm 0.002  $          & $ 1818.05    \pm 0.01  $           & 95000     & Confirmed blend \\
              & $ 0.742 \pm 0.005  $          & $ 1816.23    \pm 0.02  $           & 2000      &   \\
120362128     & $ 3.286 \pm 0.002  $          & $ 1684.425   \pm 0.01  $           & 33000     &  Candidate multiple system  \\
              & $ -                 $         & $ 1701.275   \pm 0.02  $           & 12000     &    \\
              & $ -                 $         & $ 1702.09    \pm 0.02  $           & 36000     &    \\
121945407     & $  0.9056768 \pm 0.00000002$   & $-1948.76377 \pm  0.0000001$      & 2500          & Confirmed multiple system $^{(\mathrm{a})}$  \\
              & $ 45.4711 \pm  0.00002$        & $-1500.0038  \pm  0.0004 $       & 7500          &  \\
122275115     & $ - $                         & $ 1821.779   \pm 0.01  $           & 155000    &  Candidate multiple system  \\
              & $ - $                         & $ 1830.628   \pm 0.01  $           & 63000     &  \\
              & $ - $                         & $ 1838.505   \pm 0.01  $           & 123000    &  \\
229804573     & $1.4641 \pm 0.0005$           & $ 1326.135   \pm 0.005$            & 180000    & Candidate multiple system  \\
              & $0.5283 \pm 0.0001$           & $ 1378.114   \pm 0.005$            & 9000      &   \\
252403752     & $ - $                         & $ 1817.73    \pm 0.01 $            & 2800      & Candidate multiple system \\
              & $ - $                         & $ 1829.76    \pm 0.01 $            & 23000     & \\
              & $ - $                         & $ 1833.63    \pm 0.01 $            & 5500      & \\
258837989     & $0.8870 \pm 0.001$            & $ 1599.350   \pm 0.005$            & 64000     & Candidate multiple system  \\
              & $3.0730 \pm 0.001$            & $ 1598.430   \pm 0.005$            & 25000     &   \\
266958963     & $1.5753 \pm 0.0002$           & $ 1816.425   \pm 0.001$            & 265000    & Candidate multiple system  \\
              & $2.3685 \pm 0.0001$           & $ 1817.790   \pm 0.001$            & 75000     &   \\
278956474     & $5.488068 \pm 0.000016  $     & $ 1355.400   \pm 0.005$            & 93900     & Confirmed multiple system $^{(\mathrm{b})}$ \\
              & $5.674256 \pm  -0.000030$     & $ 1330.690   \pm 0.005$            & 30000     &   \\
284925600     & $ 1.24571 \pm 0.00001 $       & $ 1765.248   \pm 0.005 $           & 490000    & Confirmed blend \\
              & $  0.31828 \pm 0.00001  $     & $ 1764.75    \pm 0.005 $           & 35000     &  \\
293954660     & $2.814 \pm  0.001 $           & $ 1739.177   \pm 0.03  $           & 272000    & Confirmed blend \\
              & $4.904 \pm 0.03 $             & $ 1739.73    \pm 0.01  $           & 9500      &  \\
312353805     & $4.951 \pm  0.003 $           & $ 1817.73    \pm 0.01  $           & 66000     & Confirmed blend \\
              & $12.89 \pm   0.01  $          & $ 1822.28    \pm 0.01$             & 19000     &  \\
318210930     & $ 1.3055432 \pm 0.000000033$  & $ -653.21602   \pm 0.0000013$      & 570000    & Confirmed multiple system $^{(\mathrm{c})}$ \\
              & $ 0.22771622 \pm 0.0000000035$& $ -732.071119 \pm 0.00000026 $     & 220000    &  \\
336434532     & $ 3.888 \pm 0.002  $          & $ 1713.66    \pm 0.01  $           & 22900     & Confirmed blend \\
              & $  0.949 \pm 0.003 $          & $ 1712.81    \pm 0.01  $           & 2900      &  \\
350622185     & $1.1686 \pm 0.0001$           & $ 1326.140   \pm 0.005$            & 200000    & Candidate multiple system  \\
              & $5.2410 \pm 0.0005$           & $ 1326.885   \pm 0.05$             & 4000      &  \\
375422201     & $9.9649 \pm 0.001$            & $ 1711.937   \pm 0.005$            & 245000    & Candidate multiple system  \\
              & $4.0750 \pm 0.001$            & $ 1713.210   \pm 0.01 $            & 39000     &  \\
376606423     & $ 0.8547 \pm 0.0002  $        & $ 1900.766   \pm 0.005 $           & 9700      & Candidate multiple system  \\
              & $  -  $                       & $ 1908.085   \pm 0.01  $           & 33000     &  \\
394177355     & $ 94.22454  \pm 0.00040 $     & $  -  $                            &  -        & Confirmed multiple system $^{(\mathrm{d})}$ \\
              & $ 8.6530941 \pm 0.0000016$    & $-2038.99492 \pm 0.00017 $         &  140000   &  \\
              & $ 1.5222468 \pm 0.0000025$    & $ -2039.1201 \pm 0.0014  $         &  -        &  \\
              & $ 1.43420486 \pm 0.00000012 $ & $-2039.23941 \pm 0.00007 $         &  -        &  \\
424508303     & $ 2.0832649 \pm 0.0000029  $ & $-3144.8661  \pm 0.0034  $          &  430000   &  Confirmed multiple system $^{(\mathrm{e})}$ \\
              & $ 1.4200401  \pm 0.0000042 $  & $-3142.5639  \pm 0.0054  $         &  250000         &  \\
441794509     & $ 4.6687 \pm  0.0002       $  & $ 1958.895   \pm 0.005   $         & 34000     &  Candidate multiple system  \\
              & $ 14.785 \pm  0.002       $   & $ 1960.845   \pm 0.005   $         & 17000     &  \\
470710327     & $ 9.9733 \pm 0.0001 $         & $ 1766.27    \pm 0.005  $          & 51000     &  Confirmed multiple system $^{(\mathrm{f})}$ \\
              & $ 1.104686 \pm 0.00001 $      & $ 1785.53266 \pm 0.000005$         & 42000     &  \\
\hline
\end{tabular}
}
\caption{
Note -- $^{(\mathrm{a})}$ KOI-6139, \citet{Borkovits2013};  
$^{(\mathrm{b})}$ \citet{2020Rowden}
$^{(\mathrm{c})}$ \citet{Koo2014}; 
$^{(\mathrm{d})}$ KOI-3156, \citet{2017Helminiak};
$^{(\mathrm{e})}$ V994 Her; \citet{Zasche2016}; 
$^{(\mathrm{f})}$ Eisner et al. {\it in prep.}
}

\label{tab:PHT-multis}

\end{table*}

\section*{Data Availability}

All of the \tess\ data used within this article are hosted and made publicly available by the Mikulski Archive for Space Telescopes (MAST, \url{http://archive.stsci.edu/tess/}). Similarly, the Planet Hunters TESS classifications made by the citizen scientists can be found on the Planet Hunters Analysis Database (PHAD, \url{https://mast.stsci.edu/phad/}), which is also hosted by MAST. All planet candidates and their properties presented in this article have been uploaded to the Exoplanet Follow-up Observing Program for TESS (ExoFOP-TESS, \url{ https://exofop.ipac.caltech.edu/tess/index.php}) website as community TOIs (cTOIs), under their corresponding TIC IDs. The ground-based follow-up observations of individual targets will be shared on reasonable request to the corresponding author.

The models of individual transit events and the data validation reports used for the vetting of the targets were both generated using publicly available open software codes, \pyaneti\ and {\sc latte}.

\section*{Acknowledgements}  

This project works under the in \textit{populum veritas est} philosophy, and for that reason we would like to thank all of the citizen scientists who have taken part in the Planet Hunters TESS project and enable us to find many interesting astrophysical systems. 

Some of the observations in the paper made use of the High-Resolution Imaging instruments `Alopeke and Zorro. `Alopeke and Zorro were funded by the NASA Exoplanet Exploration Program and built at the NASA Ames Research Center by Steve B. Howell, Nic Scott, Elliott P. Horch, and Emmett Quigley. `Alopeke and Zorro were mounted on the Gemini North and South telescope of the international Gemini Observatory, a program of NSF's NOIRLab, which is managed by the Association of Universities for Research in Astronomy (AURA) under a cooperative agreement with the National Science Foundation on behalf of the Gemini partnership: the National Science Foundation (United States), National Research Council (Canada), Agencia Nacional de Investigaci\'{o}n y Desarrollo (Chile), Ministerio de Ciencia, Tecnolog\'{i}a e Innovaci\'{o}n (Argentina), Minist\'{e}rio da Ci\^{e}ncia, Tecnologia, Inova\c{c}\~{o}es e Comunica\c{c}\~{o}es (Brazil), and Korea Astronomy and Space Science Institute (Republic of Korea). The  authors also acknowledge the very significant cultural role and sacred nature of  Maunakea. We are most fortunate to  have  the opportunity to conduct observations from this mountain.

This project has also received funding from the European Union's Horizon 2020 research and innovation programme under grant agreement N$^\circ$730890. This material reflects only the authors views and the Commission is not liable for any use that may be made of the information contained therein. This work makes use of observations from the Las Cumbres Observatory global telescope network, including the NRES spectrograph and the SBIG and Sinistro photometric instruments. 

Furthermore, NLE thanks the LSSTC Data Science Fellowship Program, which is funded by LSSTC, NSF Cybertraining Grant N$^\circ$1829740, the Brinson Foundation, and the Moore Foundation; her participation in the program has benefited this work. Finally, CJ acknowledges funding from the European Research Council (ERC) under the European Union's Horizon 2020 research and innovation programme (grant agreement N$^\circ$670519: MAMSIE), and from the Research Foundation Flanders (FWO) under grant agreement G0A2917N (BlackGEM). 

This research made use of Astropy, a community-developed core Python package for Astronomy \citep{astropy2013}, matplotlib \citep{matplotlib}, pandas \citep{pandas}, NumPy \citep{numpy}, astroquery \citep{ginsburg2019astroquery} and sklearn \citep{pedregosa2011scikit}.  




\bibliographystyle{mnras}
\bibliography{bibs} 




\appendix
\section{Planet candidate descriptions}
\label{appendixA}

A short outline all of the planet candidates, and any conclusions drawn from follow-up observations (where available). A more in depth description of the ground-based data will be presented in a follow-up paper. Unless stated otherwise, these candidates are not TOIs at the time of writing. Candidates for which we have no additional information to complement the results presented in Table~\ref{tab:PHT-caniddates} are not discussed further here.

\subsection{Single-transit planet candidates}


\textbf{TIC 103633672.} Single transit event identified in Sector 20. The single LCO/NRES spectra shows no sign of this being a double lined spectroscopic binary.  We caution that there is a star on the same pixels, which is 0.1 mag brighter. We are unable to rule this star out as the cause for the transit-like signal.

\textbf{TIC 110996418.} Single transit event identified in Sector 10. We caution that there is a star on the same \tess\ pixel, which is 2.4 magnitudes fainter than the target.

\textbf{TIC 128703021.} Single transit event identified in Sector 11. With a stellar radius of 1.6 $R_{\odot}$ and a T$_{eff}$ of 6281 this host star is likely in the subgiant phase of its evolution. The 43 spectra obtained with MINERVA australis and the two obtained with LCO/NRES are consistent with a planetary nature. Gemini speckle interferometry shows no nearby companion stars.


\textbf{TIC 142087638.} Single transit event identified in Sector 7. The best fit \pyaneti\ model of the transit suggests an orbital period of only 3.14 d. As there are no additional transits seen in the light curve, this period is clearly not possible. We caution that the transit is most likely caused by a grazing object, and is therefore likely to be caused by a stellar companion. However, without further data we are unable to rule this candidate out as being planetary in nature.

\textbf{TIC 159159904.} Single transit event identified in Sector 22. The initial two observations obtained using LCO/NRES show no sign of the candidate being a double lined spectroscopic binary.


\textbf{TIC 166184426.} Single transit event identified in Sector 11. Since the PHT discovery this cTOI has been become the priority 1 (1=highest, 5=lowest) target TOI  1955.01.

\textbf{TIC 172370679.} Single transit event identified in Sector 15. \textcolor{black}{This candidate was independently discovered and verified using a BLS algorithm used to search for transiting planets around M-dwarfs. The candidate is now the confirmed planet TOI 1899 b \citep{canas2020}.} 

\textbf{TIC 174302697.} Single transit event identified in Sectors 16. With a stellar radius of 1.6 $R_{\odot}$ and a T$_{eff}$ of 6750 this host star is likely in the subgiant phase of its evolution. \textcolor{black}{This candidate was initially flagged as a TCE and but was erroneously discounted due to the pipeline mistaking the data glitch at the time of a momentum dump as a secondary eclipse.} Since the PHT discovery this cTOI has become the priority 3 target TOI 1896.01. 

\textbf{TIC 192415680.} Single transit event identified in Sector 18. The two epochs of RV measurement obtained with OHP/SOPHIE are consistent with a planetary scenario.

\textbf{TIC 192790476.} Single transit event identified in Sector 5. This target has been identified to be a wide binary with am angular separation of 72.40 arcseconds \citep{andrews2017wideBinary} and a period of 162705 years \citep{benavides2010new}. The star exhibits large scale variability on the order of around 10 d. The signal is consistent with that of spot modulations, which would suggest that this is a slowly rotating star.



\textbf{TIC 219466784.} Single transit event identified in Sector 22. We caution that there is a nearby companion located within the same \tess\ pixel at an angular separation of 16.3 with a Vmag of 16.3". Since the PHT discovery this cTOI has become the priority 2 target TOI  2007.01.


\textbf{TIC 229055790.} Single transit event identified in Sector 21. We note that the midpoint of the transit-like events coincides with a \tess\ momentum dump, however, we believe the shape to be convincing enough to warrant further investigation. The two LCO/NRES spectra show no sign of this being a spectroscopic binary.

\textbf{TIC 229608594.} Single transit event identified in Sector 24. Since the PHT discovery this cTOI has become the priority 3 target TOI 2298.01.

\textbf{TIC 233194447.} Single transit event identified in Sector 14. The transit-like event is shallow and asymmetric and we cannot definitively rule out systematics as the cause for the event without additional data. The initial two LCO/NRES spectra show no sign of this target being a spectroscopic binary.

\textbf{TIC 237201858.} Single transit event identified in Sector 18. The single LCO/NRES spectra shows no sign of this being a double lined spectroscopic binary.

\textbf{TIC 243187830.} Single transit event identified in Sector 18. There are no nearby bright stars. \textcolor{black}{This light curve was initially flagged as a TCE, however, the flagged events corresponded to stellar variability and not the same event identified by PHT.} The single LCO/NRES spectrum shows no sign of this being a double lined spectroscopic binary. Since the PHT discovery this cTOI has become the priority 3 target TOI 2009.01.

\textbf{TIC 243417115.} Single transit event identified in Sector 11. We note that the best fit \pyaneti\ model of the transit suggests an orbital period of only 1.81 d. As there are no additional transits seen in the light curve, this period is clearly not possible. We caution that the transit is most likely caused by a grazing object, and is therefore likely to be caused by a stellar companion. However, without further follow-up data we are unable to rule this candidate out as being planetary in nature.


\textbf{TIC 264544388.} Single transit event identified in Sector 19. The single LCO/NRES spectra shows no sign of this being a double lined spectroscopic binary. Apart from the single transit event, the light curve shows no obvious signals. A periodogram of the light curve, however, reveals a series of five significant peaks, nearly equidistantly spaced by $\sim1.03$~d$^{-1}$. Additionally, a rotationally split quintuplet is visible at 7.34~d$^{-1}$, with a splitting of $\sim0.12$~d$^{-1}$, suggesting an $\ell=2$ p-mode pulsation. The Maelstrom code \citep{hey2020maelstrom} revealed pulsation timing variations which are consistent with a long period planet. \textcolor{black}{The short period signal, which was also identified by the periodogram, was flagged as a TCE, however, the single-transit event was not flagged as a TCE.} Since the PHT discovery this cTOI has become the priority 3 target TOI 1893.01.

\textbf{TIC 264766922.} Single transit event identified in Sector 8. With a stellar radius of 1.7 $R_{\odot}$ and a T$_{eff}$ of 6913 K this host star is likely entering the subgiant phase of its evolution. The V-shape of this transit and the resultant high impact parameter suggests that the object is grazing. We can therefore not rule out that this candidate it a grazing eclipsing binary.  There are clear p-mode pulsations at frequencies of 9.01 and 11.47 cycles per day, as well as possible g-mode pulsations. \textcolor{black}{A very short period signal within this light curve was flagged as a TCE, however, the single transit event was ignored by the pipeline.}

\textbf{TIC 26547036.} Single transit event identified in Sector 14. The four LCO/NRES observations are consistent with the target being a planetary body and show no sign of the signal being caused by a spectroscopic binary. We caution that there is a star on the same \tess\ pixel, however, this star is 8.2 magnitudes fainter than the target, and therefore unable to be responsible for the transit event seen in the light curve. Gemini speckle interferometry reveal no additional nearby companion stars. \textcolor{black}{This candidate was initially flagged as a TCE, however, in addition to the single transit event the pipeline identified further periodic signals that correspond to times of momentum dumps. Due to this, the candidate was never promoted to TOI status.}


\textbf{TIC 278990954.} Single transit event identified in Sector 12. With a stellar radius of 2.6 $R_{\odot}$ and a T$_{eff}$ of 5761 K this host star is likely in the subgiant phase of its evolution. We note that there are two additional stars on the same pixel as TIC 278990954. These two stars are 2.7 and 3.7 magnitudes fainter in the v-band than the target and can't be ruled out as the cause for the transit-like event without additional follow-up data.

\textbf{TIC 280865159.} Single transit event identified in Sector 16. Gemini speckle interferometry revealed any nearby companion stars. Since the PHT discovery this cTOI has become the priority 3 target TOI 1894.01.

\textbf{TIC 284361752.} Single transit event identified in Sector 26. Since the PHT discovery this cTOI has become the priority 2 target TOI 2294.01.

\textbf{TIC 296737508.} Single transit event identified in Sector 8. The single LCO/NRES and the single MINERVA australis spectra show no sign of this being a spectroscopic binary. The Sinistro snapshot image revealed no additional nearby companions.

\textbf{TIC 298663873.} Single transit event identified in Sector 19. The two LCO/NRES spectra show no sign of this being a spectroscopic binary. With a stellar radius of 1.6 $R_{\odot}$ and a T$_{eff}$ of 6750 this host star is likely in the subgiant phase of its evolution. Gemini speckle images obtained by other teams show no signs of there being nearby companion stars. Since the PHT discovery this cTOI has become the priority 3 target TOI 2180.01.

\textbf{TIC 303050301.} Single transit event identified in Sector 2. The variability of the light curve is consistent with spot modulation. A single LCO/NRES spectrum shows no signs of this being a double lined spectroscopic binary.

\textbf{TIC 303317324.} Single transit event identified in Sector 2. We note that a second transit was later seen in Sector 29, however, as this work only covers sectors 1-26 of the primary \tess\ mission, this candidates is considered a single-transit event in this work. 


\textbf{TIC 304142124.} Single transit event identified in Sector 10.\textcolor{black}{This target was independently identified as part of the Planet Finder Spectrograph, which uses precision RVs \citep{diaz2020}. This candidate is know the confirmed planet HD 95338 b.}






\textbf{TIC 331644554.} Single transit event identified in Sector 16. There is a clear mono-periodic signal in the periodogram at around 11.2 cycles per day, which is consistent with p-mode pulsation.

\textbf{TIC 332657786.} Single transit event identified in Sector 8. We caution that there is a star on the adjacent \tess\ pixel that is brighter in the V-band by 2.4 magnitudes. At this point we are unable to rule out this star as the cause of the transit-like signal. 


\textbf{TIC 356700488.} Single transit event identified in Sector 16. There is a clear mono-periodic signal in the periodogram at around 1.2 cycles per day, which is consistent with either spot modulation or g-mode pulsation. However, there is no clear signal visible in the light curve that would allow us to differentiate between these two scenarios based on the morphology of the variation. Since the PHT discovery this cTOI has become the priority 3 target TOI 2098.01.

\textbf{TIC 356710041.} Single transit event identified in Sector 23. With a stellar radius of 2.8 $R_{\odot}$ and a T$_{eff}$ of 5701 K this host star is likely in the subgiant phase of its evolution. \textcolor{black}{This candidate was initially flagged as a TCE, however, in addition to the single transit event, the pipeline identified a further event that corresponds to the time of a momentum dump. Due to this the candidate failed the `odd-even test' and was initially discarded as a TOI.} Since the PHT discovery this cTOI has become the priority 3 target TOI 2065.01

\textbf{TIC 369532319.} Single transit event identified in Sector 16. Gemini speckle interferometry revealed no nearby companion stars.


\textbf{TIC 384159646.} Single transit event identified in Sector 12. The eight LCO/NRES and six MINERVA australis spectra are consistent with this candidate being a planet. Both the SBIG snapshot and the Gemini speckle interferometry observations revealed no companion stars. Since the PHT discovery this cTOI has become the priority 3 target TOI 1895.01.



\textbf{TIC 418255064.} Single transit event identified in Sector 12. The Gemini speckle image shows no sign of nearby companions.


\textbf{TIC 422914082.} Single transit event identified in Sector 4. Single Sinistro snapshot image reveals no additional nearby stars.

\textbf{TIC 427344083.} Single transit event identified in Sector 24. We note that there is a star on the adjacent \tess\ pixel to the target, which is 3.5 magnitude fainter in the V-band than the target star. We also caution that the V-shape of the transit and the high impact parameter suggest that this is a grazing transit. However, without additional follow-up observations we are unable to rule this candidate out as a planet.

\textbf{TIC 436873727.} Single transit event identified in Sector 18. The host star shows strong variability on the order of one day, which is consistent with spot modulations or g-mode pulsations. The periodogram reveals multi-periodic behaviour in the low frequency range consistent with g-mode pulsations. 

\textbf{TIC 452920657.} Single transit event identified in Sector 17. The V-shape of this transit suggests that the object is grazing and future follow-up observations may reveal this to be an EB. \textcolor{black}{This candidate was initially flagged as a TCE, however, in addition to the single transit event, the pipeline identified a further two event that corresponds to likely stellar variability. Due to this the candidate failed the `odd-even test' and was initially discarded as a TOI.}

\textbf{TIC 455737331.} Single transit event identified in Sector 17. We note that there is a star on the same \tess\ pixel as the target, which is 4.5 magnitudes fainter in the V-band. Neither the SBIG snapshot nor the Gemini speckle interferometry revealed any further nearby companion stars.

\textbf{TIC 456909420.} Single transit event identified in Sector 17. We caution that the V-shape of the transit and the high impact parameter suggest that this is a grazing transit. However, without additional follow-up observations we are unable to rule this candidate out as a planet.



\textbf{TIC 53843023.} Single transit event identified in Sector 1. We caution that the high impact parameter returned by the best fit \pyaneti\ model suggests that the transit event is caused by a grazing body. However, at this point we are unable to rule this candidate out as being planetary in nature.

\textbf{TIC 63698669.} Single transit event identified in Sector 2. The SBIG snapshot image revealed no nearby companions. \textcolor{black}{This candidate was initially identified as a TCE, however, in addition to the single transit event, the pipeline identified a further 3 events the light curve. Due to these, additional events, which correspond to stellar variability, the candidate was not initially promoted to TOI status.} However, since the PHT discovery this cTOI has become TOI 1892.01.

\textbf{TIC 70887357.} Single transit event identified in Sector 5. With a stellar radius of 2.1 $R_{\odot}$ and a T$_{eff}$ of 5463 K this host star is likely in the subgiant phase of its evolution. \textcolor{black}{This candidate was initially flagged as a TCE, however, in addition to the single transit-event the pipeline identified a further signal, and thus failed the `odd-even' transit test.} However, since the PHT discovery this cTOI has become the priority 3 target TOI 2008.01.



\textbf{TIC 91987762.} Single transit event identified in Sector 21. \textcolor{black}{This candidate was initially flagged as a TCE, however, in addition to the single transit-event the pipeline identified a further signal, and thus failed the `odd-even' transit test.} Since the PHT discovery this cTOI has become the priority 3 target TOI 1898.01.


\subsection{Multi-transit and multi-planet candidates}

\textbf{TIC 160039081.} Multi-transit candidate with a period of 30.2 d. Single LCO/NRES spectra shows no sign of this being a double lined spectroscopic binary and a snapshot image using SBIG shows no nearby companions. The Gemini speckle images also show no additional nearby companions. Since the PHT discovery this cTOI has  become the priority 1 target TOI 2082.01.

\textbf{TIC 167661160.} Multi-transit candidate with a period of 36.8 d. The nine LCO/NRES and four MINERVA australis spectra have revealed this to be a long period eclipsing binary.

\textbf{TIC 179582003.} Multi-transit candidate with a period of 104.6 d. There is a clear mono-periodic signal in the periodogram at around 0.59 cycles per day, which is consistent with either spot modulation or g-mode pulsation. We caution that this candidate is located in a crowded field. With a stellar radius of 2.0 $R_{\odot}$ and a T$_{eff}$ of 6115 K this host star is likely in the subgiant phase of its evolution.

\textbf{TIC 219501568.} Multi-transit candidate with a period of 16.6 d. With a stellar radius of 1.7 $R_{\odot}$ and a T$_{eff}$ of 6690 K this host star is likely entering the subgiant phase of its evolution. \textcolor{black}{This candidate was identified as a TCE, however, it was not initially promoted to TOI status as the signal was thought to be off-target by the automated pipeline.} However, since the PHT discovery this cTOI has become the priority 3 target TOI 2259.01

\textbf{TIC 229742722.} Multi-transit candidate with a period of 63.48 d. Eight LCO/NRES and four OHP/SOPHIE observations are consistent with this candidate being a planet. Gemini speckle interferometry reveals no nearby companion stars. \textcolor{black}{This candidate was flagged as a TCE in sector 20, where it only exhibits a single transit event. An additional event was identified at the time of a momentum dump, and as such it failed the `odd-even' test and was not initially promoted to TOI status.} However, since the PHT discovery this cTOI has become the priority 3 target TOI 1895.01.

\textbf{TIC 235943205.} Multi-transit candidate with a period of 121.3 d. The LCO/NRES and OHP/SOPHIE observations remain consistent with a planetary nature of the signal. Since the PHT discovery this cTOI has become the priority 3 target TOI 2264.01.

\textbf{TIC 267542728.} Multi-transit event with period of 39.7 d. Observations obtained with Keck showed that the RV shifts are not consistent with a planetary body and are most likely due to an M-dwarf companion.

\textbf{TIC 274599700.} Multi-transit candidate with a period of 33.0 d. One of the two transit-like even is only half visible, with the other half of the event falling in a \tess\ data gap.



\textbf{TIC 328933398.} Multi-planet candidates. The 2-minute cadence light curve shows two single transit events of different depths across two \tess\ sectors, both of which are consistent with an independent planetary body. In addition to the short cadence data, this target was observed in an additional three sectors as part of the 30-minute cadence full frame images. These showed additional transit events for one of the planet candidates, with a period of 24.9 d. \textcolor{black}{This light curve was initially flagged as containing a TCE event, however, the two 2-minute cadence single transit events were thought to belong to the same transiting planet. The TCE was initially discarded as the pipeline identified the events to be off-target.} However, since the PHT discovery these two cTOIs has become the priority 3 and 1 targets, TOI 1873.01 and TOI 1873.01, respectively.

\textbf{TIC 349488688.} Multi-planet candidate, with one single transit event and one multi-transit candidate with a period of 11 d. Two LCO/NRES and two OHP/SOPHIE spectra, along with ongoing HARPS North are consistent with both of these candidates being planetary in nature. \textcolor{black}{The single transit event was initially identified as a TCE, however, in addition to the event it identified two other signals at the time of momentum dumps, and was therefore initially discarded by the pipeline as it failed the `odd-even' transit test.} However, since the PHT discovery the two-transit event has become the 1 targets, TOI 2319.01 (Eisner et al. in prep).

\textbf{TIC 385557214.} Multi-transit candidate with a period of 5.6 d. The prominent stellar variation seen in the light curve is likely due to spots or pulsation The high impact parameter returned by the best fit \pyaneti\ modelling suggests that the transit is likely caused by a grazing object. Without further observations, however, we are unable to rule this candidate out as being planetary in nature. \textcolor{black}{This candidate was flagged as a TCE but was not promoted to TOI status due to the other nearby stars.}

\textbf{TIC 408636441.} Multi-transit candidate with a period of 18.8 or 37.7 d. Due to \tess\ data gaps, half of the period stated in Table~\ref{tab:PHT-caniddates} is likely. The SBIG snapshot and Gemini speckle images show no signs of companion stars. \textcolor{black}{This candidate was flagged as a TCE in sector 24, where it only exhibits a single transit event. An additional event was identified at the time of a momentum dump, and as such it failed the `odd-even' test and was not initially promoted to TOI status.} However, since the PHT discovery this cTOI has become the priority 3 target TOI 1759.01.

\textbf{TIC 441642457.} Multi-transit candidate with a period of 79.8 d. \textcolor{black}{This candidate was flagged as a TCE in sector 14, where it only exhibits a single transit event. An additional event was identified at the time of a momentum dump, and as such it failed the `odd-even' test and was not initially promoted to TOI status.} Since the PHT discovery this cTOI has become the priority 2 target TOI 2073.01.

\textbf{TIC 441765914.} Multi-transit candidate with a period of 161.6 d.  Since the PHT discovery this cTOI has become the priority 1 target TOI 2088.01.

\textbf{TIC 48018596.} Multi-transit candidate with a period of 100.1 days (or a multiple thereof). The single LCO/NRES spectrum shows no sign of this target being a double lined spectroscopic binary. Gemini speckle interferometry revealed no nearby companion stars.  \textcolor{black}{This candidate was initially flagged as a TCE, however, in addition to the transit-events, the pipeline classified, what we consider stellar variability as an additional event. As such it failed the `odd-even' transit test and wasn't promoted to TOI status.} However, since the PHT discovery this cTOI has become the priority 3 target TOI 2295.01.

\textbf{TIC 55525572.} Multi-transit candidate with a period of 83.9 d. Since the PHT discovery this cTOI has become the confirmed planet TOI 813 \citep{2020eisner}.

\textbf{TIC 82452140.} Multi-transit candidate with a period of 21.1 d. Since the PHT discovery this cTOI has become the priority 2 target TOI 2289.01.



\label{lastpage}
\end{document}